\documentclass[10pt,a4paper]{article}
\usepackage[dvips]{color}
\usepackage{epsfig}
\usepackage{epstopdf}
\usepackage{amsmath}
\usepackage{graphicx}
\usepackage{amssymb,amsmath}
\usepackage{amsmath}
\usepackage{tabularx}

\begin{document}
\title{\bf On cosmology of interacting varying polytropic dark fluids}
\author{{Martiros 
Khurshudyan$^{a,b,c}$\thanks{Email:khurshudyan@yandex.ru, 
khurshudyan@tusur.ru}~~, Asatur 
Khurshudyan$^{d}$\thanks{Email:khurshudyan@mechins.sci.am}}\\
$^{a}${\small {\em International Laboratory for Theoretical Cosmology, 
TUSUR, Tomsk, Russia}}\\
$^{b}${\small {\em Research Division, TSPU, 
Tomsk, Russia}}\\
$^{c}${\small {\em Institute of Physics, University of Zielona Gora, 
Zielona Gora, Poland}}\\
$^{d}${\small {\em Institute of Mechanics, National Academy of Sciences 
of Armenia, Yerevan, Armenia}}\\
}\maketitle

\begin{abstract}
In this paper a possibility of the accelerated expansion of the large 
scale universe with a varying polytropic fluid of a certain type is 
presented. About a special role of non-gravitational interactions 
between dark energy and dark matter, in particular, about a possibility 
of improvement and solution of problems arising in modern cosmology, has 
been discussed for a long time. This motivates us to consider new 
models, where non-gravitational interactions between varying polytropic 
fluid and cold dark matter are allowed. Mainly non-linear interactions 
of a specific type is considered, found in recent literature. In order 
to understand the behavior of suggested cosmological models, besides 
cosmographic analysis, $Om$ analysis is applied. Moreover, with 
different datasets, including a strong gravitational lensing dataset, 
the observational constraints on the model parameters are obtained using 
$\chi^{2}$ analysis.

\end{abstract}

\section{Introduction}\label{sec:INT}
Available observational data indicating an accelerated expanding flat 
universe suggests an existence of dark energy and dark matter, if the 
dynamics of the background is according to general relativity~\cite{PC}~(and references therien to follow how the accelerated expansion of the universe have been detected for the first time). However, 
recent research on this problem indicates alternative scenarios as well 
including modification of general relativity~\cite{M1}~-~\cite{M8} and, for instance, 
gravitationally induced particle creation~\cite{P1}~-~\cite{P7}. The aim of all developed 
approaches is to generate an appropriate negative pressure to cancel the 
attractive nature of gravity. It is clear, that the source of 
anti-gravity cannot be arbitrary and cannot destroy the recent universe 
and existing symmetries. Moreover, it should have appropriate properties 
not to alter the dynamics started from the birth of the universe. 
Therefore, suggested models should pass astrophysical and cosmological 
tests and the models of dark energy and dark matter should be 
constrained from available observational data. On the other hand, 
possible tension existing between different observational datasets from 
one hand side, and the technological limitations on the other hand, 
allows a scanning of the physics of our universe up to some redshifts. 
Incompleteness of this kind makes impossible to finalize our knowledge 
giving the final models for dark energy and dark matter. Therefore, a 
phenomenological approach to parameterize the darkness of the recent 
universe in a form of dark energy and dark matter in recent literature 
has been used frequently. In particular, recently, a phenomenological 
modification of polytropic dark fluid has been suggested and the study 
showed that the model with the following equation of state~\cite{Me1}
\begin{equation}
P = -A H^{-k} \rho^{u},
\end{equation}
where $A$, $k$ and $u$ are constants, while $H$ is the Hubble parameter, 
can be used as a source of anti-gravity. $A$, $k$ and $u$ parameters 
should be determined from the observational data. The interest towards 
to polytropic type fluids is related to their applications in 
astrophysics. On the other hand, in recent literature alternative models 
of dark fluids like Chaplygin gas and various viscous dark fluids, also 
able to solve the problem, systematically have been presented~\cite{Me2}~-~~\cite{Me4}~(and references therein). In 
general, dark energy can be parameterized via the equation of state, 
which provides a functional dependency of the pressure from the energy 
density: the examples are Chaplygin gas and polytropic fluid with their different modifications. The second option 
includes a parameterization of the energy density of the source and the 
examples are ghost dark energy and generalized holographic dark energy 
models with Nojiri-Odintsof cut-offs~(Ricci dark energy and other 
holographic dark energy models are particular examples of the 
holographic dark energy with Nojiri-Odintsov cut-offs)~\cite{Me5}~-~\cite{Me15}. Finally, the 
third option describing dark energy can be a parameterization of the 
equation of state parameter of dark energy~(see for instance \cite{S1}). It is well known that the 
simplest model of dark energy is the cosmological constant $\Lambda$, 
which provides results in well correspondence with available 
observational data. However, there are two main problems that 
$\Lambda$CDM faced and then the need of introducing of dynamical dark 
energy models raised. The first attempt to build a dynamical dark energy 
model has been based on the idea of varying cosmological constant 
$\Lambda(t)$. Various phenomenological models of $\Lambda(t)$ have been 
considered in literature successfully and there is a significant attempt 
to use of a false vacuum decay to construct models of $\Lambda(t)$. 
Sometimes in literature such models represented as the models of dark 
energy based on quantum theory~(see for instance~\cite{U1}~-~\cite{U3} and reference therein for more information). Other models of dynamical dark 
energy, besides mentioned dark fluid models, are quintessence, phantom, 
quintom and k-essence scalar field dark energy models among the others~\cite{DE1}. 
On the other, hand mentioned dark energy models are introduced by hand 
in to the dynamics of the background, therefore a direct modification of 
general relativity accounts as a straightforward way to explain the 
accelerated expansion of the large scale universe. Moreover, a 
modification of general relativity has proved to be useful also for the 
physics of the early universe including a possibility to explain the 
inflation~\cite{In1}~-~\cite{In2}~(and references therein). In this paper, we already had mentioned that general 
relativity will describe the background dynamics and it is known that 
with such models it is very useful to use the idea of non-gravitational 
interaction~(we refer the readers to the cited papers for more information on this idea). Usually, it is accounted that the non-gravitational 
interaction is a specific type of interaction which is deduced from the 
properties of dark energy and dark matter. Therefore, in cosmological 
models we usually consider interaction between dark energy and dark 
matter only. Non-gravitational interaction is not only a 
phenomenological assumption, it also allows to improve theoretical 
results, therefore there is also an increasing amount of interest 
towards this option/idea. It is believed, that with the understanding of 
the structures of dark energy and dark matter, this question will be 
understood as well. There are also some models of non-gravitational 
interactions which are already can be considered as classical ones. 
Moreover, there are new developments in this direction mainly in recent 
literature, again based on various phenomenological modifications. In 
this paper we will study new cosmological models involving new forms of 
non-linear non-gravitational interactions considered in recent 
literature and the main goal is to demonstrate the viability of these 
models to the problems of the accelerated expansion of the recent 
universe. Moreover, we will use $Om$ analysis and the relative change of 
this parameter~\cite{Om}
\begin{equation}\label{eq:DeltaOm}
\Delta Om = 100 \times \left [ \frac{Om_{Model}}{Om_{\Lambda CDM}} - 
1\right ]
\end{equation}
to study the possible differences between the new models and 
$\Lambda$CDM model, for which $Om = \Omega^{(0)}_{dm} = 0.27$. Moreover, 
$\chi^{2}$ analysis will be used in order to constrain the parameters of 
the models.\\

The paper is organised as follows: In section~\ref{sec:CCDM} a 
description of the background dynamics with the datasets in use are 
presented. Moreover, the description of $Om$ analysis is also presented 
in section~\ref{sec:CCDM}. In section~\ref{sec:MODELS}, the best fit 
values of the parameters obtained during $\chi^{2}$ analysis for 
appropriate datasets are presented and appropriate key consequences are 
discussed for all models. Finally, discussion on obtained results and 
possible future extension of considered cosmological model are 
summarized in section~\ref{sec:Discussion}.

\section{Background dynamics and datasets}\label{sec:CCDM}

In case on interacting dark energy models we should take into account 
that the dynamics of the energy densities of dark energy and dark matter 
should be modified. In particular, the following differential equation 
should be considered
\begin{equation}\label{eq:DM}
\dot{\rho}_{dm} + 3 H \rho_{dm} = Q,
\end{equation}
and
\begin{equation}\label{eq:DE}
\dot{\rho}_{de} + 3 H (\rho_{de} + P_{de} ) = -Q,
\end{equation}
where $Q$ indicates non-gravitational interaction and dark matter is 
assumed to be cold with $P=0$. Such representation directly depends on 
the assumption that the effective fluid in the universe will be 
described as follows
\begin{equation}
P_{eff} = P_{de},
\end{equation}
\begin{equation}
\rho_{eff} = \rho_{dm} + \rho_{de}.
\end{equation}
On the other hand, in an isotropic and spatially homogeneous flat FRW c
universe, the Friedmann equations are as follows
\begin{equation}
H^{2} = \frac{8\pi G}{3} \rho_{eff},
\end{equation}
\begin{equation}
\frac{\ddot{a}}{a} = -\frac{4 \pi G}{3} ( \rho_{eff} + 3P_{eff}),
\end{equation}
and describe the background dynamics. To separate a physically 
reasonable solution in case of a phenomenological assumption it is 
necessary to constrain the parameters of the models. In this paper we 
will use the following datasets
\begin{enumerate}
\item The differential age of old galaxies, given by $H(z)$.
\item The peak position of baryonic acoustic oscillations (BAO).
\item The SN Ia data.
\item Strong Gravitation Lensing data.
\end{enumerate}
In the case of the Observed Hubble Data, one defines chi-square given by
\begin{equation}
\chi^{2}_{OHD} = \sum \frac{\left ( H(\textbf{P},z) - H_{obs}(z)\right 
)^{2}}{\sigma_{OHD}^{2}},
\end{equation}
where $H_{obs}(z)$ is the observed Hubble parameter at redshift $z$ and 
$\sigma_{OHD}$ is the error associated with that particular observation, 
while $ H(\textbf{P},z)$ is the Hubble parameter obtained from the model 
and $\textbf{P}$ is the set of the parameters to be 
determined/constrained from the dataset. On the other hand, $7$ 
measurements have been jointly used determining the BAO~(Baryon Acoustic 
Oscillation) peak parameter to constrain the models by
\begin{equation}
\chi^{2}_{BAO} = \sum \frac{\left ( A(\textbf{P},z) - A_{obs}(z)\right 
)^{2}}{\sigma_{BAO}^{2}},
\end{equation}
where the theoretical value for the $\textbf{P}$ set of the parameters 
$A(\textbf{P},z)$ is determined as
\begin{equation}
A(\textbf{P},z_{1}) = \frac{\sqrt{\Omega_{m}}}{E(z_{1})^{1/3}} \left( 
\frac{ \int_{0}^{z_{1}} \frac{dz}{E(z)} }{z_{1}}\right)^{2/3},
\end{equation}
with $E(z) = H(z)/H_{0}$ and $H_{0}$ is the value of the Hubble 
parameter at $z=0$. For the Supernovae Data, $\chi^{2}_{\mu}$ is defined 
as
\begin{equation}
\chi^{2}_{\mu} = A - \frac{B^{2}}{C},
\end{equation}
where
\begin{equation}
A = \sum {\frac{(\mu(\textbf{P},z) - \mu_{obs})^{2}}{\sigma_{\mu}^{2}}},
\end{equation}
\begin{equation}
B = \sum {\frac{\mu(\textbf{P},z) - \mu_{obs}}{\sigma_{\mu}^{2}}} ,
\end{equation}
and
\begin{equation}
C = \sum {\frac{1}{\sigma_{\mu}^{2}}}.
\end{equation}
In the last $3$ equations $\sigma_{\mu}$ is the uncertainty in the 
distance modulus~\cite{Data1}. We will follow to the receipt of Ref.~\cite{Data2} and use the 
data presented there in order to obtain constraints on the parameters of 
the models from the strong gravitational lensing. To obtain appropriate 
constraints, we will look for the set of the values of the parameters of 
the models to minimize $\chi^{2}$ function defined as
\begin{equation}
\chi^{2} = \chi^{2}_{OHD} + \chi^{2}_{BAO}+ \chi^{2}_{\mu} + 
\chi^{2}_{SGL},
\end{equation}
if we want to obtain the constraints using all datasets presented above.

\section{Models and data fitting}\label{sec:MODELS}

Three different types of models will be analyzed in this paper involving 
different forms of non-gravitational interactions between dark energy 
and dark matter. The forms of non-gravitational interactions have been considered for the first time in Ref.~\cite{Me2}. The parameters of the models to be constrained using 
$\chi^{2}$ statistical tool are as follows $\textbf{P} = \{ H_{0}, 
\Omega^{(0)}_{dm}, A, b, u, k \}$. In order to simplify the discussion 
on the results of the fit to find the best fit values of the parameters 
and the discussion on a relative change of $Om$ parameter, we organized 
appropriate subsections.

\subsection{Models of the first type}

In the case of the models consider in this subsection the following 
general form describing the non-gravitation interaction between dark 
energy and dark matter will be taken into account
\begin{equation}\label{eq:M1}
Q = 3 H b \left( \rho_{de} + \rho_{dm} + \frac{\rho_{de}^{2}}{\rho_{de} 
+ \rho_{dm}}\right),
\end{equation}
where $b$ is a constant, $H$ is the Hubble parameter, while $\rho_{de}$ 
and $\rho_{dm}$ represent the energy density of the varying polytropic 
dark fluid under the consideration and the energy density of dark 
matter, respectively. However, before the presentation of the results 
associated to this model we will study other two models as well, where 
non-gravitational interactions between dark energy and dark matter are 
particular examples obtained from Eq.~(\ref{eq:M1}).

\subsubsection{Case $1$}\label{ssec:1_1}

The study shows, that when the interaction is defined in the following 
way
\begin{equation}\label{eq:M1_1}
Q = 3 H b \left( \rho_{de} + \frac{\rho_{de}^{2}}{\rho_{de} + 
\rho_{dm}}\right),
\end{equation}
then using all datasets described above give the results presented in 
Table~\ref{tab:Table1}.
\begin{table}
   \centering
     \begin{tabular}{ | l | l | l | l | l | l | p{1cm} |}
     \hline
  $\chi^{2}$ & $\Omega^{(0)}_{DM} (f)$ & $H_{0} (f)$ & $A$ & $b$ & $u 
(f)$ & $k$ \\
       \hline
  $782.03$ & $0.27$ & $71.9$ & $0.965$ & $0.024$ & $1.0$  & $-0.014$\\
           \hline
  $781.60$ & $0.27$ & $71.9$ & $0.741$ & $0.033$ & $1.25$  & $0.463$ \\
     \hline
  $781.37$ & $0.27$ & $71.9$ & $0.879$ & $0.047$ & $1.5$  & $1.047$ \\
     \hline
  $781.23$ & $0.28$ & $71.9$ & $0.966$ & $0.033$ & $1.0$  & $-0.014$ \\
     \hline
  $780.90$ & $0.28$ & $71.9$ & $0.655$ & $0.043$ & $1.25$  & $0.437$ \\
      \hline
  $780.73$ & $0.28$ & $71.9$ & $0.879$ & $0.057$ & $1.5$  & $1.047$ \\
      \hline
  $782.56$ & $0.29$ & $71.9$ & $0.965$ & $0.043$ & $1.0$  & $-0.014$ \\
      \hline
   $782.28$ & $0.29$ & $71.9$ & $0.655$ & $0.052$ & $1.25$  & $0.437$ \\
      \hline
  $782.19$ & $0.29$ & $71.9$ & $0.503$ & $0.062$ & $1.5$  & $0.914$ \\
      \hline
   $785.81$ & $0.30$ & $71.9$ & $0.943$ & $0.052$ & $1.0$  & $-0.014$ \\
      \hline
    $785.62$ & $0.30$ & $71.9$ & $0.914$ & $0.066$ & $1.25$  & $0.516$ \\
      \hline
    $785.59$ & $0.30$ & $71.9$ & $0.542$ & $0.072$ & $1.5$  & $0.914$ \\
      \hline
    $790.91$ & $0.31$ & $71.9$ & $0.845$ & $0.062$ & $1.0$  & $-0.041$ \\
      \hline
    $790.79$ & $0.31$ & $71.9$ & $0.517$ & $0.071$ & $1.25$  & $0.384$ \\
      \hline
    $790.82$ & $0.31$ & $71.9$ & $0.502$ & $0.081$ & $1.5$  & $0.914$ \\
      \hline

     \end{tabular}
\caption{The best fit results for the model with $Q = 3 H b \left( 
\rho_{de} + \frac{\rho_{de}^{2}}{\rho_{de} + \rho_{dm}}\right)$ with 
$\chi^{2}_{OHD} + \chi^{2}_{BAO} + \chi^{2}_{SGL} + \chi^{2}_{\mu}$. $f$ 
means that the parameter has been fixed to the presented value in 
advance befor the fit has been started.}
   \label{tab:Table1}
\end{table}
The fit has been performed having the following priors on $H_{0} = 
71.9$, $A \in [-2,2]$ and $b\in [-1,1]$. The presented results in 
Table~\ref{tab:Table1} are for $\Omega^{(0)}_{dm} = 0.27$, 
$\Omega^{(0)}_{dm} = 0.28$, $\Omega^{(0)}_{dm} = 0.29$, 
$\Omega^{(0)}_{dm} = 0.30$, $\Omega^{(0)}_{dm} = 0.31$ for $u = 1$, $u = 
1.25$ and $u = 1.5$, respectively. Initial priors for $u$ was $u 
\in(0,3]$, while for $\Omega^{(0)}_{DM}$ was $\Omega^{(0)}_{DM} \in 
[0.26,0.32]$. The value of the Hubble parameter has been chosen 
according to the report of 2016 provided by Hubble Space Telescope 
mission~\cite{HST}. From the results presented in Table~\ref{tab:Table1}, we see 
that the minimum for $\chi^{2}_{OHD} + \chi^{2}_{BAO} + \chi^{2}_{SGL} + 
\chi^{2}_{\mu} = 780.73$, for instance, has been obtained for $\{ H_{0}, 
\Omega^{(0)}_{dm}, A, b, u, k \} = \{ 71.9, 0.28, 0.879, 0.057, 1.5, 
1.047 \}$ providing the best fit of theoretical model with observational 
data. We would like to mention that the best fit values of the 
parameters of the models has not been affected, when we have considered 
$\chi^{2}_{OHD} + \chi^{2}_{BAO} + \chi^{2}_{\mu}$ i.e. considered 
strong gravitational lensing data does not play an important role on the 
best fit values of the parameters.
\begin{figure}[h!]
  \begin{center}$
  \begin{array}{cccc}
\includegraphics[width=80 mm]{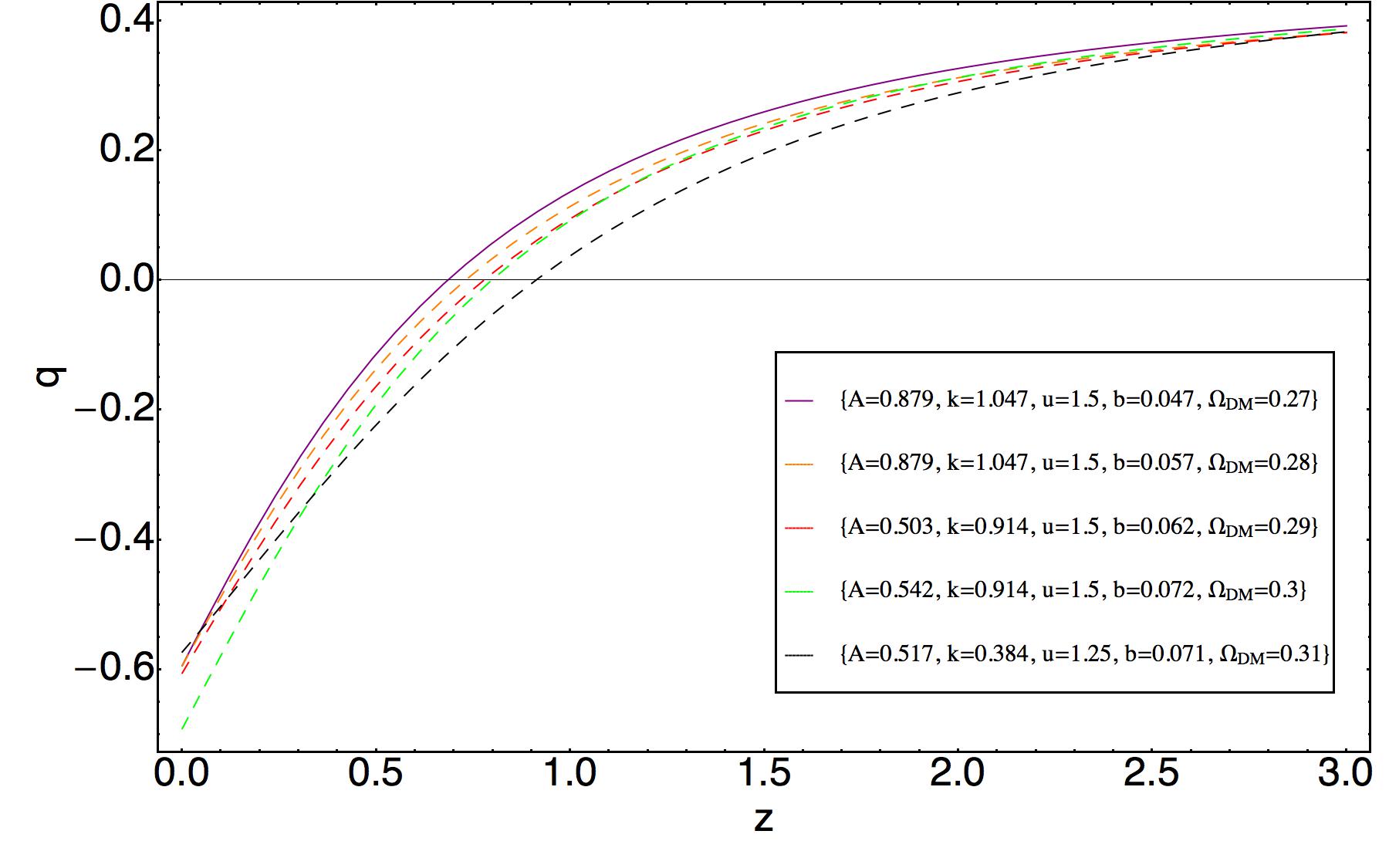} &
\includegraphics[width=80 mm]{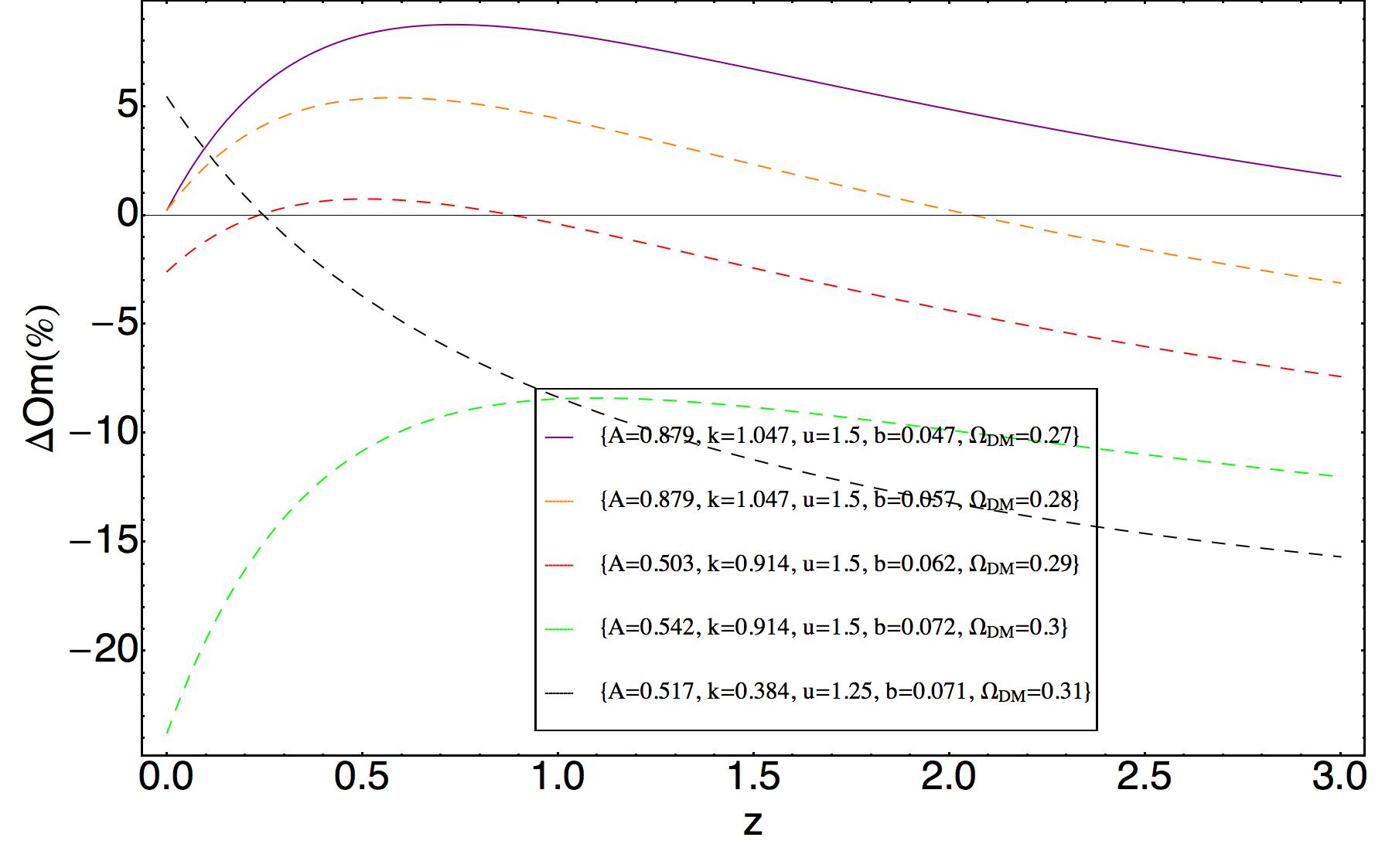}
  \end{array}$
  \end{center}
\caption{The graphical behavior of the deceleration parameter $q$ and 
$\Delta Om$, Eq.~(\ref{eq:DeltaOm}), against the redshift $z$. The 
considered model is free from the cosmological coincidence problem. The 
form of non-gravitational interaction is given by Eq.~(\ref{eq:M1_1}).}
  \label{fig:Fig1}
\end{figure}
Fig.~(\ref{fig:Fig1}) presents the graphical behavior of the 
deceleration parameter $q$ and $\Delta Om$. The behavior of the 
deceleration parameter indicates that considered model can explain the 
accelerated expansion of the universe. Moreover, a phase transition 
between decelerated expanding and accelerated expanding phases with 
different redshifts is on face. On the other hand, the graphical 
behavior of $\Delta Om$ presented on the right plot of 
Fig.~(\ref{fig:Fig1}) represents differences between the new model and 
$\Lambda$CDM standard cosmological model. In particular, the analysis 
shows that for $\Omega^{(0)}_{dm} = 0.27$ and $\Omega^{(0)}_{dm} = 0.28$ 
cases the relative change at $z=0.0$ is about $0.1 \%$. On the other 
hand, the study showed that $\{ H_{0}, \Omega^{(0)}_{dm}, A, b, u, k \} 
= \{ 71.9, 0.30, 0.542, 0.072, 1.5, 0.914 \}$ case satisfies to the 
known constraints from a modified two point $Om$ analysis with the 
result from BOSS experiment for the Hubble parameter at $z=2.34$~\cite{BOSS}. In 
this case the present value of the equation of state parameter for 
considered varying polytropic dark fluid is $\omega_{de} = -1.14$ i.e. 
in this case the phantom line crossing is possible and $\Delta Om 
\approx -24 \%$. Moreover, if we will take into account constraint om 
$\omega_{de}$ obtained from PLANCK 2015 experiment, then the 
applicability of this model will be under the doubt. On the other hand, 
the model with the following parameters $\{ H_{0}, \Omega^{(0)}_{dm}, A, 
b, u, k \} = \{ 71.9, 0.31, 0.517, 0.071, 1.25, 0.384 \}$ with 
$\omega_{de} = -1.022$ will satisfy to the mentioned 
constraints~($\Delta Om \approx 5.5 \%$). Having obtained results, we 
conclude that an additional data is needed in order to be able for any 
future conclusion.

\subsubsection{Case $2$}\label{ssec:1_2}

On the other hand, the study shows, that when the interaction $Q$ has 
the following form
\begin{equation}\label{eq:M1_2}
Q = 3 H b \left( \rho_{dm} + \frac{\rho_{de}^{2}}{\rho_{de} + 
\rho_{dm}}\right),
\end{equation}
the, for instance, the minimum $\chi^{2}_{OHD} + \chi^{2}_{BAO} + 
\chi^{2}_{SGL} + \chi^{2}_{\mu} = 780.67$ providing the best fit has 
been obtained with $\{ H_{0}, \Omega^{(0)}_{dm}, A, b, u, k \} = \{ 
71.9, 0.28, 0.948, 0.047, 1.25, 0.516 \}$. In Table~\ref{tab:Table2} a 
comprehensive information is provided allowing to understand how 
$\Omega^{(0)}_{dm}$ and $u$~(both fixed in advance before the fit has 
been started) affect on the fit results with fixed value of the Hubble 
parameter $H_{0} = 71.9$. On the other hand, imposing the constraints 
from a modified two point $Om$ analysis, the result from BOSS experiment 
for the Hubble parameter at $z=2.34$ and the constraints on the 
$\omega_{de}$ from PLANCK 2015, only one option has been survived among 
presented in Table~\ref{tab:Table2}: $\{ H_{0}, \Omega^{(0)}_{dm}, A, b, 
u, k \} = \{ 71.9, 0.28, 0.689, 0.057, 1.0, -0.094 \}$ with $\omega_{de} 
= -1.03$ and $\Delta Om \approx 3.5$ at $z=0.0$. The transition redshift 
in this case is $z_{tr} \approx 0.9$.

\begin{table}
   \centering
     \begin{tabular}{ | l | l | l | l | l | l | p{1cm} |}
     \hline
  $\chi^{2}$ & $\Omega^{(0)}_{DM} (f)$ & $H_{0} (f)$ & $A$ & $b$ & $u 
(f)$ & $k$ \\
       \hline
  $781.78$ & $0.27$ & $71.9$ & $0.983$ & $0.019$ & $1.0$  & $-0.014$\\
           \hline
  $781.37$ & $0.27$ & $71.9$ & $0.672$ & $0.033$ & $1.25$  & $0.437$ \\
     \hline
  $781.22$ & $0.27$ & $71.9$ & $0.897$ & $0.052$ & $1.5$  & $1.047$ \\
     \hline
  $780.96$ & $0.28$ & $71.9$ & $0.776$ & $0.028$ & $1.0$  & $-0.067$ \\
     \hline
  $780.67$ & $0.28$ & $71.9$ & $0.948$ & $0.047$ & $1.25$  & $0.516$ \\
      \hline
  $780.68$ & $0.28$ & $71.9$ & $0.812$ & $0.062$ & $1.5$  & $1.021$ \\
      \hline
  $782.26$ & $0.29$ & $71.9$ & $0.776$ & $0.038$ & $1.0$  & $-0.067$ \\
      \hline
   $782.12$ & $0.29$ & $71.9$ & $0.948$ & $0.057$ & $1.25$  & $0.516$ \\
      \hline
  $782.22$ & $0.29$ & $71.9$ & $0.914$ & $0.071$ & $1.5$  & $1.047$ \\
      \hline
   $785.55$ & $0.30$ & $71.9$ & $0.689$ & $0.047$ & $1.0$  & $-0.094$ \\
      \hline
    $785.54$ & $0.30$ & $71.9$ & $0.534$ & $0.062$ & $1.25$  & $0.384$ \\
      \hline
    $785.74$ & $0.30$ & $71.9$ & $0.517$ & $0.076$ & $1.5$  & $0.914$ \\
      \hline
    $790.72$ & $0.31$ & $71.9$ & $0.689$ & $0.057$ & $1.0$  & $-0.094$ \\
      \hline
    $790.83$ & $0.31$ & $71.9$ & $0.534$ & $0.071$ & $1.25$  & $0.384$ \\
      \hline
    $791.13$ & $0.31$ & $71.9$ & $0.517$ & $0.085$ & $1.5$  & $0.914$ \\
      \hline

     \end{tabular}
\caption{The best fit results for the model with $Q = 3 H b \left( 
\rho_{dm} + \frac{\rho_{de}^{2}}{\rho_{de} + \rho_{dm}}\right)$ with 
$\chi^{2}_{OHD} + \chi^{2}_{BAO} + \chi^{2}_{SGL} + \chi^{2}_{\mu}$. $f$ 
means that the parameter has been fixed to the presented value in 
advance befor the fit has been started.}
   \label{tab:Table2}
\end{table}

\subsubsection{Case $3$}

The model considered in this subsection admits the interaction between 
dark energy and dark matter given by Eq.~(\ref{eq:M1}). Taking into 
account the same priors as it was in two other cases considered in 
\ref{ssec:1_1} and \ref{ssec:1_2}, we found the best fit of theoretical 
results with  $\{ H_{0}, \Omega^{(0)}_{dm}, A, b, u, k \} = \{ 71.9, 
0.27, 0.569, 0.024, 1.5, 0.943 \}$ giving  $\chi^{2} = \chi^{2}_{OHD} + 
\chi^{2}_{SGL} + \chi^{2}_{\mu} = 778.18$. The constraining of the 
parameters of the model with $\chi^{2} = \chi^{2}_{OHD} + \chi^{2}_{SGL} 
+ \chi^{2}_{\mu} + \chi^{2}_{BAO}$ reveals that the best fit does not 
affected. On the other, when we used only $H(z)$ data, then the best fit 
has been obtained with $\chi^{2}_{OHD} = 15.81$ and $\{ H_{0}, 
\Omega^{(0)}_{dm}, A, b, u, k \} = \{ 71.9, 0.3, 0.948, 0.037, 1.5, 
1.021 \}$. The result presented here do not satisfy the constraints from 
the modified two point $Om$ analysis, the result from BOSS experiment 
for the Hubble parameter at $z=2.34$ and the constraints on the 
$\omega_{de}$ from PLANCK 2015. Therefore, it is important to study the 
model in the light of strong gravitational lensing to improve the best 
fit values and see how the new results change the situation with 
mentioned constraints.

\subsection{Models of the second type}

The main model to be studied in this subsection admits the following 
form of non-gravitational interaction between varying polytropic dark 
energy and cold dark matter
\begin{equation}\label{eq:M2_3}
Q = 3 H b \left( \rho_{de} + \rho_{dm} + 
\frac{\rho_{de}\rho_{dm}}{\rho_{de} + \rho_{dm}}\right),
\end{equation}
where $b$ is the parameter and should be determined from observational 
data under consideration. Before to present the main results obtained 
for this model, we will discuss other models of non-gravitational 
interactions, which are particular examples of more general form 
presented by Eq.~(\ref{eq:M2_3}).

\subsubsection{Case $1$}

The comparison of theoretical results with observational data reveals 
that when the non-gravitational interaction is given as follows
\begin{equation}\label{eq:M2_1}
Q = 3 H b \left( \rho_{de} + \frac{\rho_{de}\rho_{dm}}{\rho_{de} + 
\rho_{dm}}\right),
\end{equation}
and when $H_{0} = 71.9$, $\Omega^{(0)}_{dm} = 0.27$ and $u = 1.5$ are 
fixed, then with minimum $\chi^{2}_{OHD} + \chi^{2}_{BAO} + 
\chi^{2}_{SGL} + \chi^{2}_{\mu} = 781.30$ the best fit will be obtained 
providing the following values $A = 0.879$, $b = 0.052$ and $k = 1.044$ 
for the rest of the parameters. On the other hand, the best fit values 
of the parameters had been obtained for $H_{0} = 71.9$, 
$\Omega^{(0)}_{dm} = 0.28$ and $u = 1.5$ with $A = 0.810$, $b = 0.062$ 
and $k = 1.025$ giving the following minimum value $\chi^{2}_{OHD} + 
\chi^{2}_{BAO} + \chi^{2}_{SGL} + \chi^{2}_{\mu} = 780.71$. Moreover, in 
case of $H_{0} = 71.9$, $\Omega^{(0)}_{dm} = 0.29$ and $u = 1.5$ with 
$\chi^{2}_{OHD} + \chi^{2}_{BAO} + \chi^{2}_{SGL} + \chi^{2}_{\mu} = 
782.23$ providing the best fit has been found giving $A$, $b$ and $k$ to 
be $0.689$, $0.071$ and $0.987$, respectively. Finally, the best fit 
values of the parameters of the model have been found when 
$\Omega^{(0)}_{dm} = 0.30$ and $\Omega^{(0)}_{dm} = 0.31$ have been 
fixed in advance giving $\{A, b, u, k \} = \{ 0.931, 0.071, 1.25, 0.516 
\}$ and $\{A, b, u, k \} = \{ 0.931, 0.081, 1.25, 0.516 \}$, 
respectively~(the parameter $u$ and $H_{0}$ have been fixed in advance 
as well). On the other hand, a joint constraint from the modified two 
point $Om$ analysis, the result from BOSS experiment for the Hubble 
parameter at $z=2.34$ and the constraints on the $\omega_{de}$ from 
PLANCK 2015 imply that $\{H_{0}, \Omega^{(0)}_{dm}, A, b, u, k \} = \{ 
71.9, 0.30, 0.931, 0.071, 1.25, 0.516 \}$ with $\omega_{de} = -1.047$ is 
the best result among obtained once. In this case the transition 
redshift $z_{tr} \approx 0.85$ and $\Delta Om = -1.0 \%$.

\subsubsection{Case $2$}

\begin{figure}[h!]
  \begin{center}$
  \begin{array}{cccc}
\includegraphics[width=80 mm]{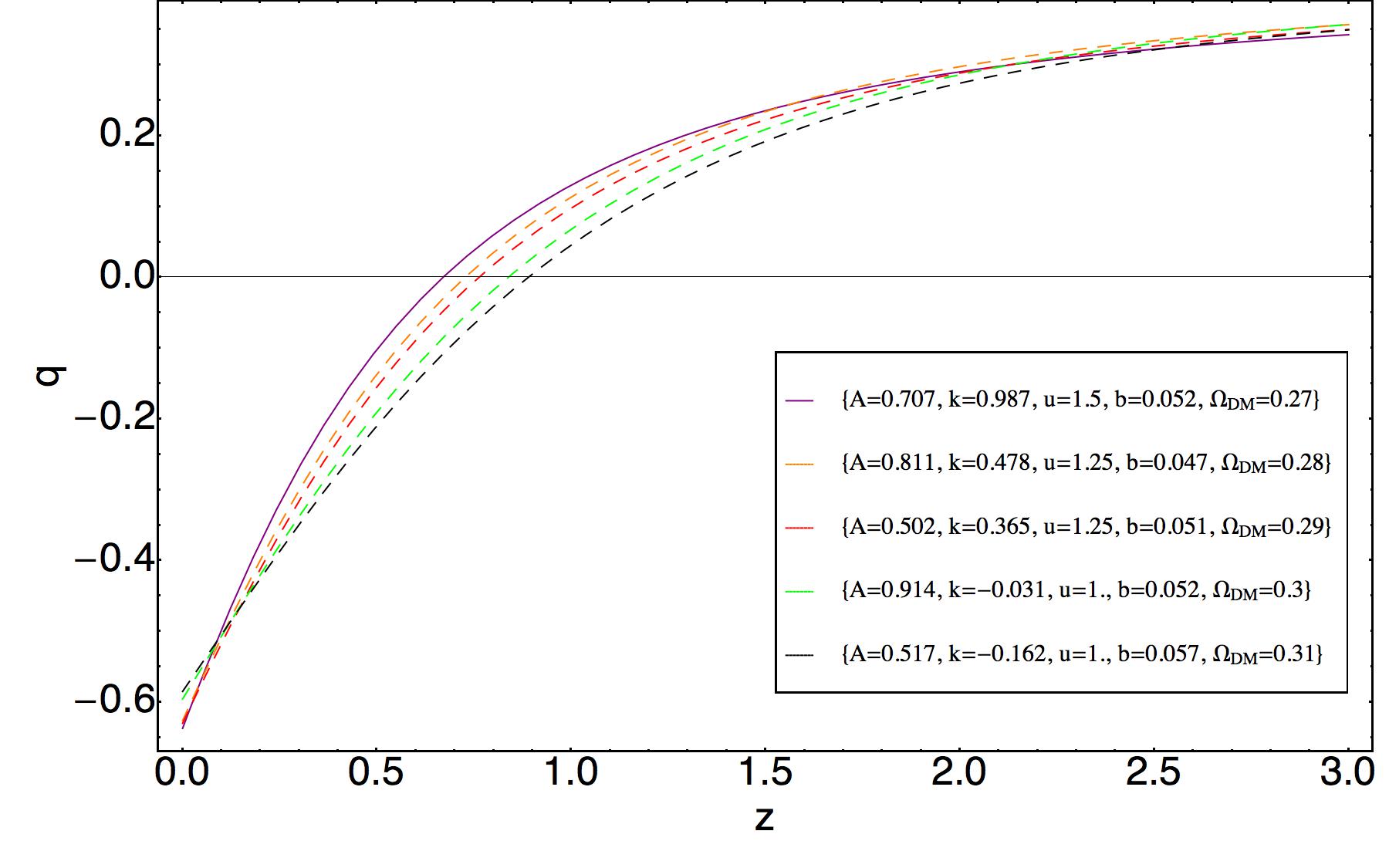} &
\includegraphics[width=80 mm]{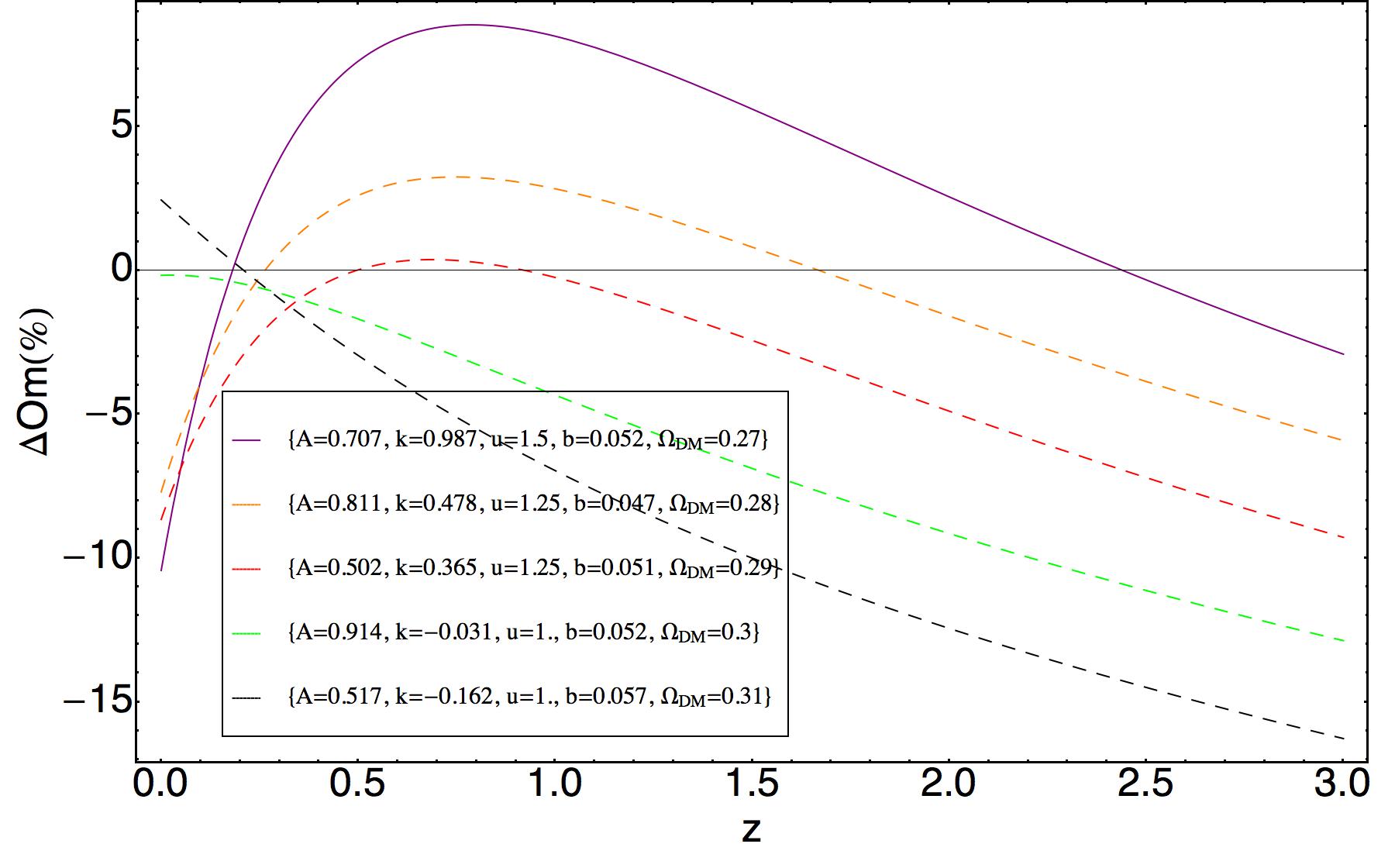}
  \end{array}$
  \end{center}
\caption{The graphical behavior of the deceleration parameter $q$ and 
$\Delta Om$, Eq.~(\ref{eq:DeltaOm}), against the redshift $z$. The 
considered model is free from the cosmological coincidence problem. The 
form of non-gravitational interaction is given by Eq.~(\ref{eq:M2_2}).}
  \label{fig:Fig2}
\end{figure}

When $H_{0} = 71.9$, $\Omega^{(0)}_{dm} = 0.27$ and $u = 1.5$ are fixed 
in advance, then with minimum $\chi^{2}_{OHD} + \chi^{2}_{BAO} + 
\chi^{2}_{SGL} + \chi^{2}_{\mu} = 781.22$ the best fit will be obtained 
when $A = 0.707$, $b = 0.052$ and $k = 0.987$ for the model described by 
the interaction of the following form
\begin{equation}\label{eq:M2_2}
Q = 3 H b \left( \rho_{dm} +  \frac{\rho_{de}\rho_{dm}}{\rho_{de} + 
\rho_{dm}}\right)
\end{equation}
On the other hand, for the same model with fixed $\Omega^{(0)}_{dm} = 
0.28$, the following result have been obtained $\{A, b, u, k \} = \{ 
0.811, 0.047, 1.25, 0.478 \}$ giving $\chi^{2}_{OHD} + \chi^{2}_{BAO} + 
\chi^{2}_{SGL} + \chi^{2}_{\mu} = 780.65$ minimal value. The results 
corresponding to fixed $\Omega^{(0)}_{dm} = 0.29, 0.30, 0.31$ with 
appropriate minimal value of $\chi^{2}_{OHD} + \chi^{2}_{BAO} + 
\chi^{2}_{SGL} + \chi^{2}_{\mu}$ are presented in 
Table~\ref{tab:Table3}. On the left plot of Fig.~(\ref{fig:Fig2}) the 
graphical behavior of the deceleration parameter $q$ is presented 
indicating the accelerated expansion of the large scale universe. On the 
other hand, the right plot represents the graphical behavior of $\Delta 
Om$. Similar to previously discussed cases additional constraints 
indicates, that in future we should take the following $\{H_{0}, 
\Omega^{(0)}_{dm}, A, b, u, k \} = \{ 71.9, 0.30, 0.914, 0.052, 1.0, 
-0.031 \}$ constraints providing the best fit of the theoretical results 
with considered observational data. In this case the transition redshift 
is $z_{tr} \approx 0.85$, while $\Delta Om \approx -1.0 \%$ and 
$\omega_{de} = -1.044$.

\begin{table}
   \centering
     \begin{tabular}{ | l | l | l | l | l | l | p{1cm} |}
     \hline
  $\chi^{2}$ & $\Omega^{(0)}_{DM} (f)$ & $H_{0} (f)$ & $A$ & $b$ & $u 
(f)$ & $k$ \\
      \hline
  $782.16$ & $0.29$ & $71.9$ & $0.502$ & $0.051$ & $1.25$  & $0.365$ \\
      \hline
   $785.56$ & $0.30$ & $71.9$ & $0.914$ & $0.052$ & $1.0$  & $-0.031$ \\
      \hline
    $790.79$ & $0.31$ & $71.9$ & $0.517$ & $0.057$ & $1.0$  & $-0.162$ \\
      \hline
     \end{tabular}
\caption{The best fit results for the model with $Q = 3 H b \left( 
\rho_{dm} +  \frac{\rho_{de}\rho_{dm}}{\rho_{de} + \rho_{dm}}\right)$ 
with $\chi^{2}_{OHD} + \chi^{2}_{BAO} + \chi^{2}_{SGL} + 
\chi^{2}_{\mu}$. $f$ means that the parameter has been fixed to the 
presented value in advance before the fit has been started.}
   \label{tab:Table3}
\end{table}

\subsubsection{Case $3$}

\begin{figure}[h!]
  \begin{center}$
  \begin{array}{cccc}
\includegraphics[width=80 mm]{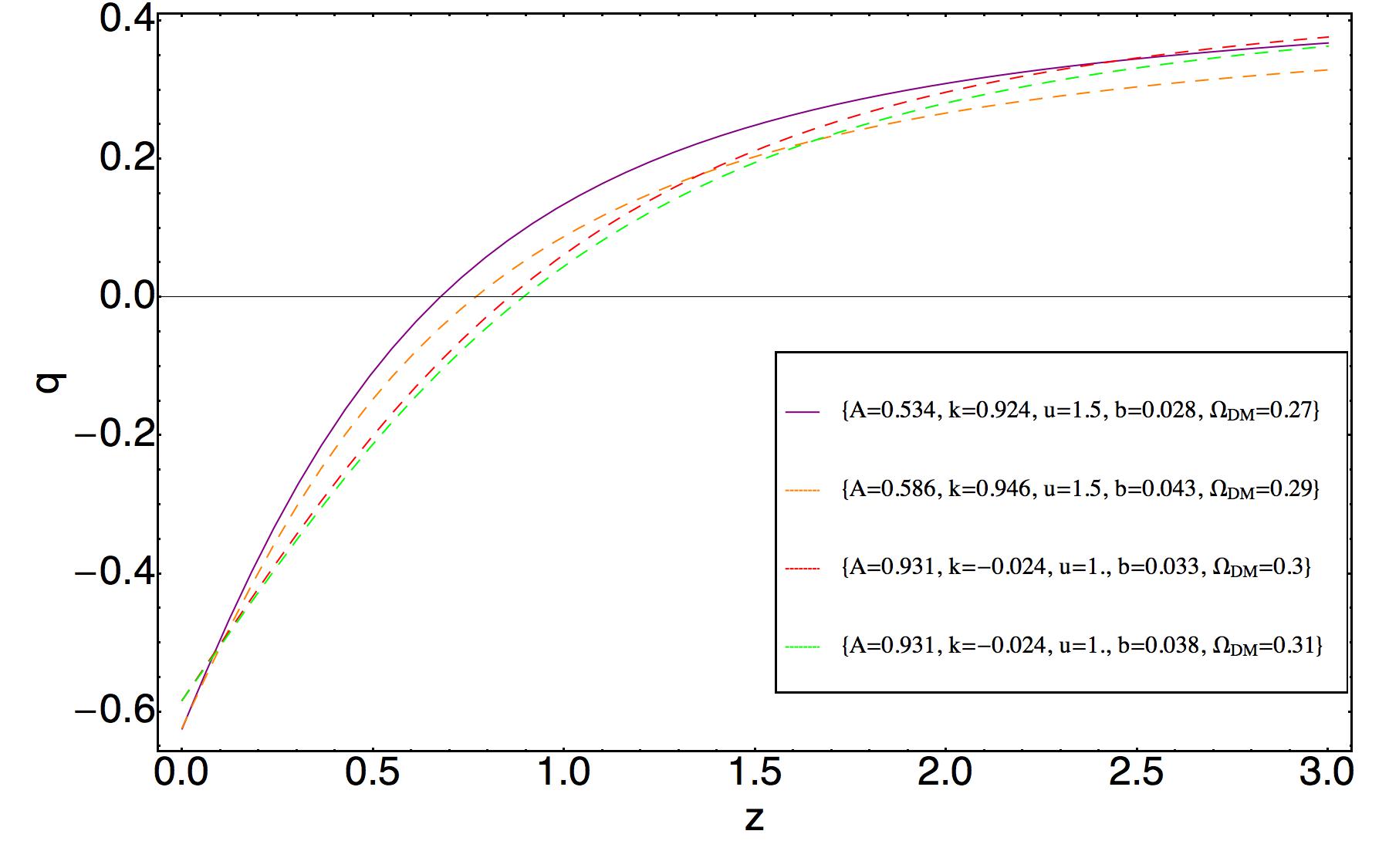} &
\includegraphics[width=80 mm]{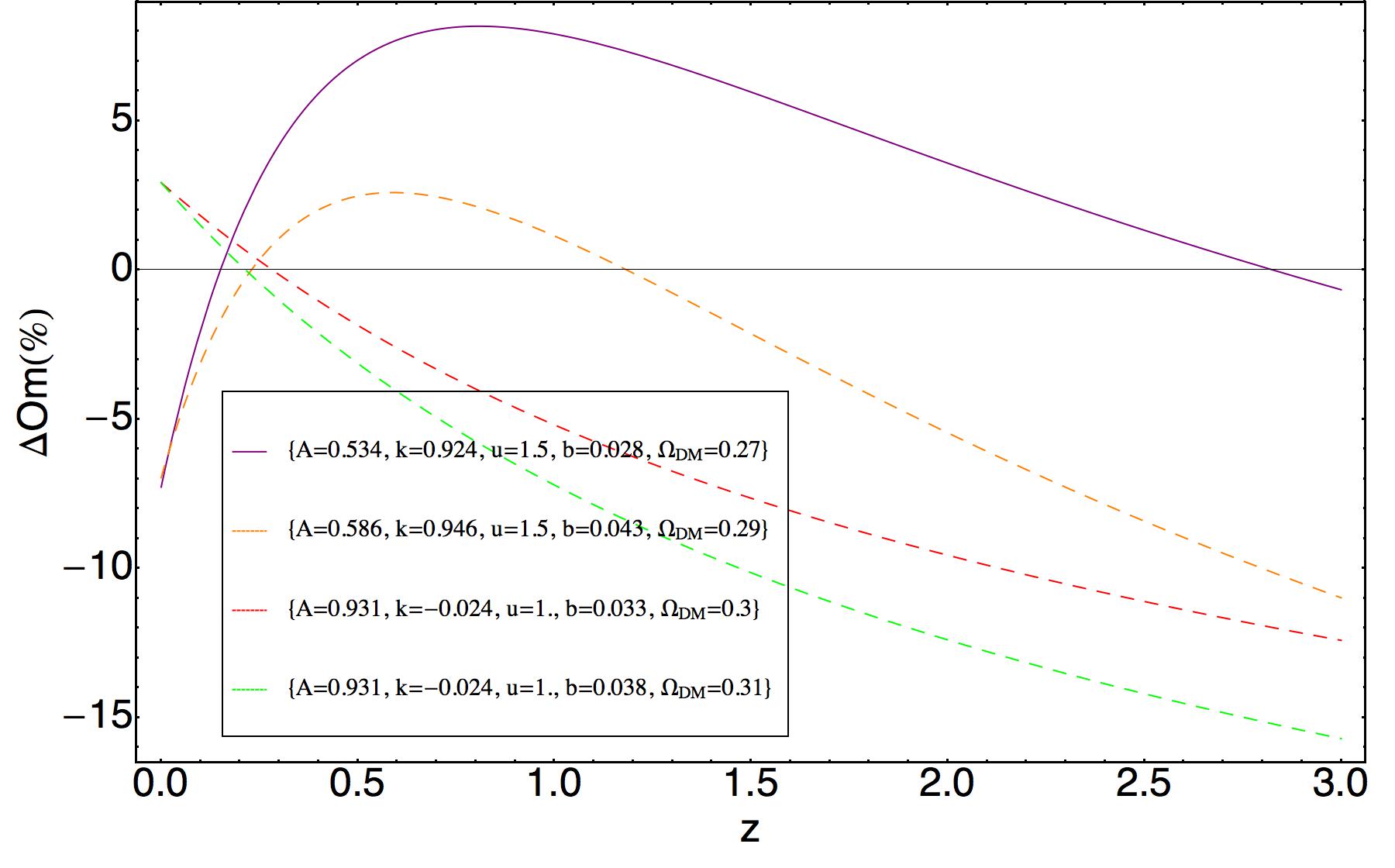} \\
\includegraphics[width=80 mm]{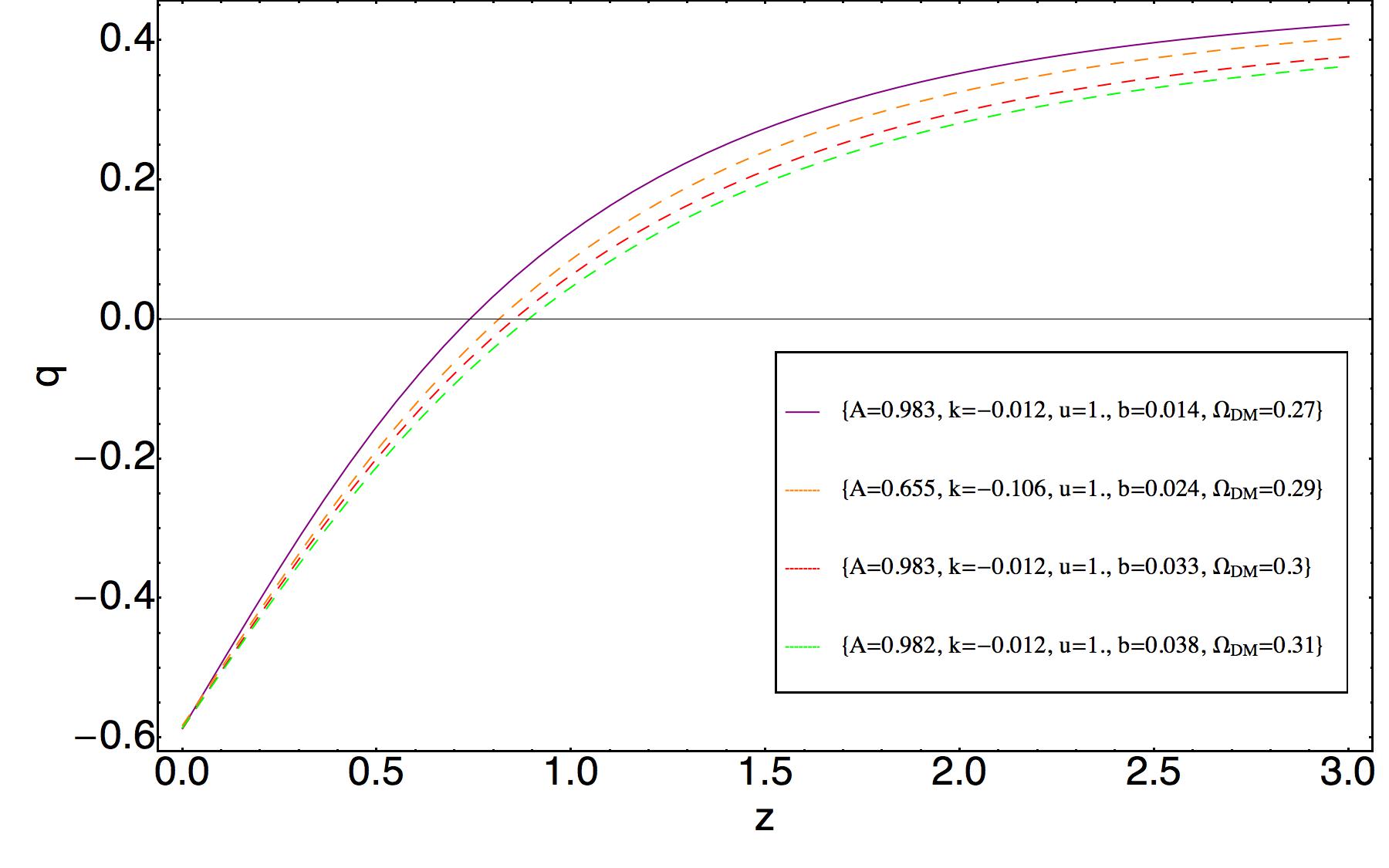} &
\includegraphics[width=80 mm]{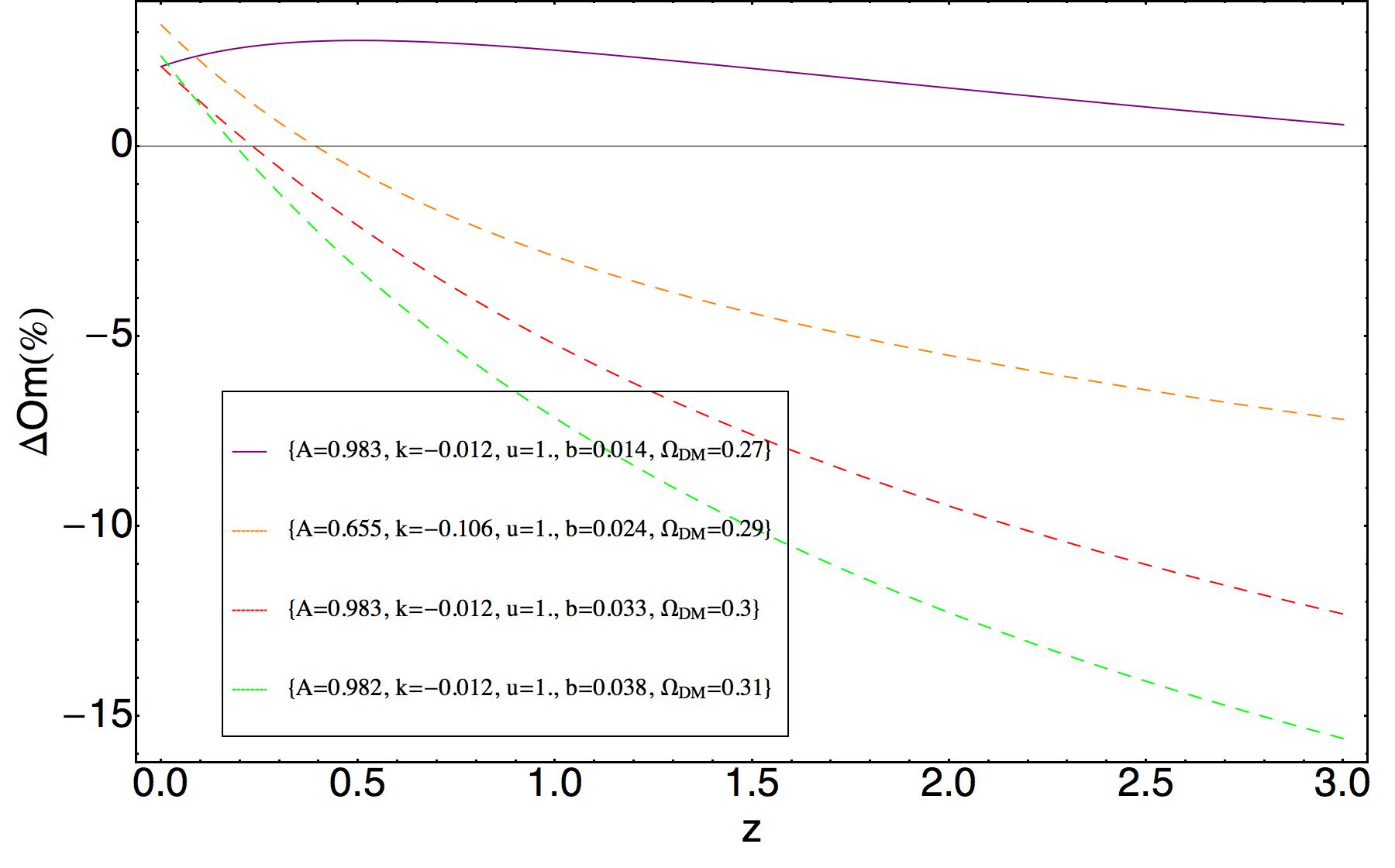} \\
  \end{array}$
  \end{center}
\caption{The graphical behavior of the deceleration parameter $q$ and 
$\Delta Om$, Eq.~(\ref{eq:DeltaOm}), against the redshift $z$. The 
considered model is free from the cosmological coincidence problem. The 
form of non-gravitational interaction is given by Eq.~(\ref{eq:M2_3}). 
The top panel represents the result corresponding to the analysis with 
$\chi^{2}_{OHD} + \chi^{2}_{BAO} + \chi^{2}_{\mu}$, while the bottom 
panel represents the results for the analysis with $\chi^{2}_{OHD} + 
\chi^{2}_{BAO} + \chi^{2}_{SGL} + \chi^{2}_{\mu}$.}
  \label{fig:Fig3}
\end{figure}

When the non-gravitational interaction is given by Eq.~(\ref{eq:M2_3}), 
then the best fit values of the parameters of the model with $H_{0} = 
71.9$, $\Omega^{(0)}_{DM} = 0.27$ and $u = 1.5$ fixed in advance are as 
follows: $\{A, b, k \} = \{ 0.534, 0.028, 0.924 \}$ giving 
$\chi^{2}_{OHD} + \chi^{2}_{BAO} + \chi^{2}_{\mu} = 563.29$. On the 
other hand, with $H_{0} = 71.9$, $\Omega^{(0)}_{DM} = 0.29$ and $u = 
1.5$ fixed in advance the best fit values of the other parameters of the 
model are as follows: $\{A, b, k \} = \{ 0.586, 0.043, 0.946 \}$ giving 
$\chi^{2}_{OHD} + \chi^{2}_{BAO} + \chi^{2}_{\mu} = 564.37$. Moreover, 
with $\Omega^{(0)}_{DM} = 0.30$ and $\Omega^{(0)}_{DM} = 0.31$ fixed in 
advance the minimal values of $\chi^{2}_{OHD} + \chi^{2}_{BAO} + 
\chi^{2}_{\mu}$ with $567.70$ and $572.89$ provided the best fit of 
theoretical results with observational data when  $\{A, b, u, k \} = \{ 
0.931, 0.033, 1.0, -0.024 \}$ and $\{A, b, u, k \} = \{ 0.931, 0.038, 
1.0, -0.024 \}$, respectively~($u$ also had been fixed in advance). The 
results corresponding to $\chi^{2}_{OHD} + \chi^{2}_{BAO} + 
\chi^{2}_{SGL} + \chi^{2}_{\mu}$ are presented in 
Table~\ref{tab:Table4}. The top panel of Fig.~(\ref{fig:Fig2}) 
represents the behavior of the deceleration parameter $q$ and $\Delta 
Om$ corresponding to the best fit values obtained for the parameters of 
the model using the data from the differential age of old galaxies, 
given by $H(z)$, the peak position of baryonic acoustic oscillations 
(BAO) and the SN Ia data, when $\Omega^{(0)}_{DM}$, $u$ and $H_{0}$ were 
fixed in advance. On the other hand the bottom panel of 
Fig.~(\ref{fig:Fig3}) represents the graphical behavior of the same 
parameters with the same parameters of the model fixed in advance, when 
together with mentioned observational datasets the strong gravitational 
lensing data has been used. After imposing the constraints as has been 
discussed for the other cases we found the following picture
\begin{enumerate}
\item in case of the analysis with $\chi^{2}_{OHD} + \chi^{2}_{BAO} + 
\chi^{2}_{\mu}$ we should take $\{A, b, u, k \} = \{ 0.931, 0.033, 1.0, 
-0.024 \}$ for the candidate supported from considered constraints
\item in case of the analysis with $\chi^{2}_{OHD} + \chi^{2}_{BAO} + 
\chi^{2}_{SGL} + \chi^{2}_{\mu}$ the results presented in 
Table~\ref{tab:Table4} for $\Omega^{(0)}_{DM} = 0.29, 0.30, 0.31$ are 
the candidates supported from considered constraints. However, only the 
result corresponding to $\Omega^{(0)}_{DM} = 0.30$ will be accounted as 
a candidate providing the best fit.
\end{enumerate}

\begin{table}
   \centering
     \begin{tabular}{ | l | l | l | l | l | l | p{1cm} |}
     \hline
  $\chi^{2}$ & $\Omega^{(0)}_{DM} (f)$ & $H_{0} (f)$ & $A$ & $b$ & $u 
(f)$ & $k$ \\
      \hline
  $781.79$ & $0.27$ & $71.9$ & $0.983$ & $0.014$ & $1.0$  & $-0.012$ \\
      \hline
  $782.35$ & $0.29$ & $71.9$ & $0.655$ & $0.024$ & $1.0$  & $-0.106$ \\
      \hline
  $785.62$ & $0.30$ & $71.9$ & $0.983$ & $0.033$ & $1.0$  & $-0.012$ \\
      \hline
  $790.78$ & $0.31$ & $71.9$ & $0.983$ & $0.038$ & $1.0$  & $-0.012$ \\
      \hline

     \end{tabular}
\caption{The best fit results for the model with $Q = 3 H b \left( 
\rho_{de} + \rho_{dm} + \frac{\rho_{de}\rho_{dm}}{\rho_{de} + 
\rho_{dm}}\right)$ with $\chi^{2}_{OHD} + \chi^{2}_{BAO} + 
\chi^{2}_{SGL} + \chi^{2}_{\mu}$. $f$ means that the parameter has been 
fixed to the presented value in advance before the fit has been 
started.}
   \label{tab:Table4}
\end{table}

\subsection{Models of the third type}

\begin{figure}[h!]
  \begin{center}$
  \begin{array}{cccc}
\includegraphics[width=80 mm]{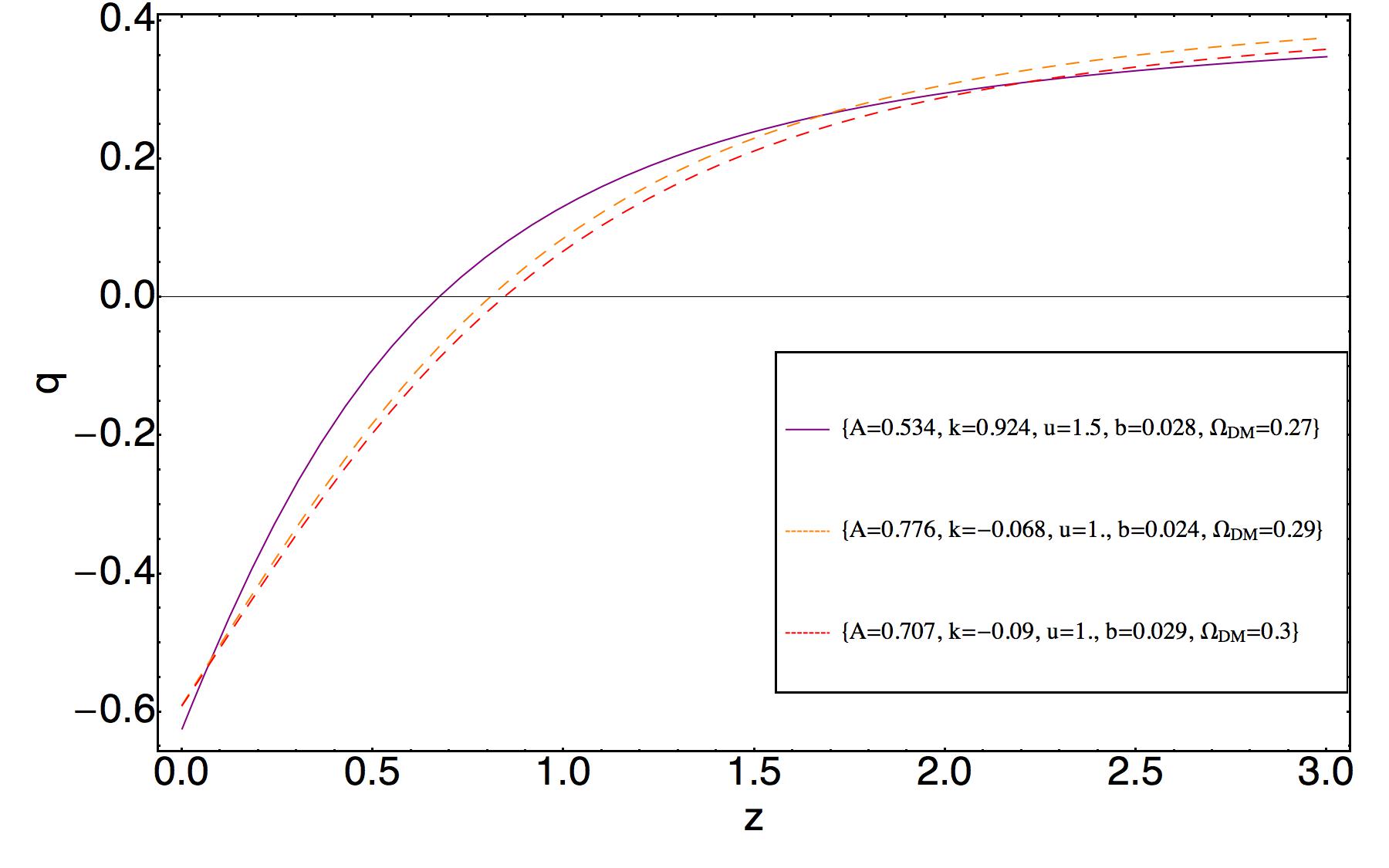} &
\includegraphics[width=80 mm]{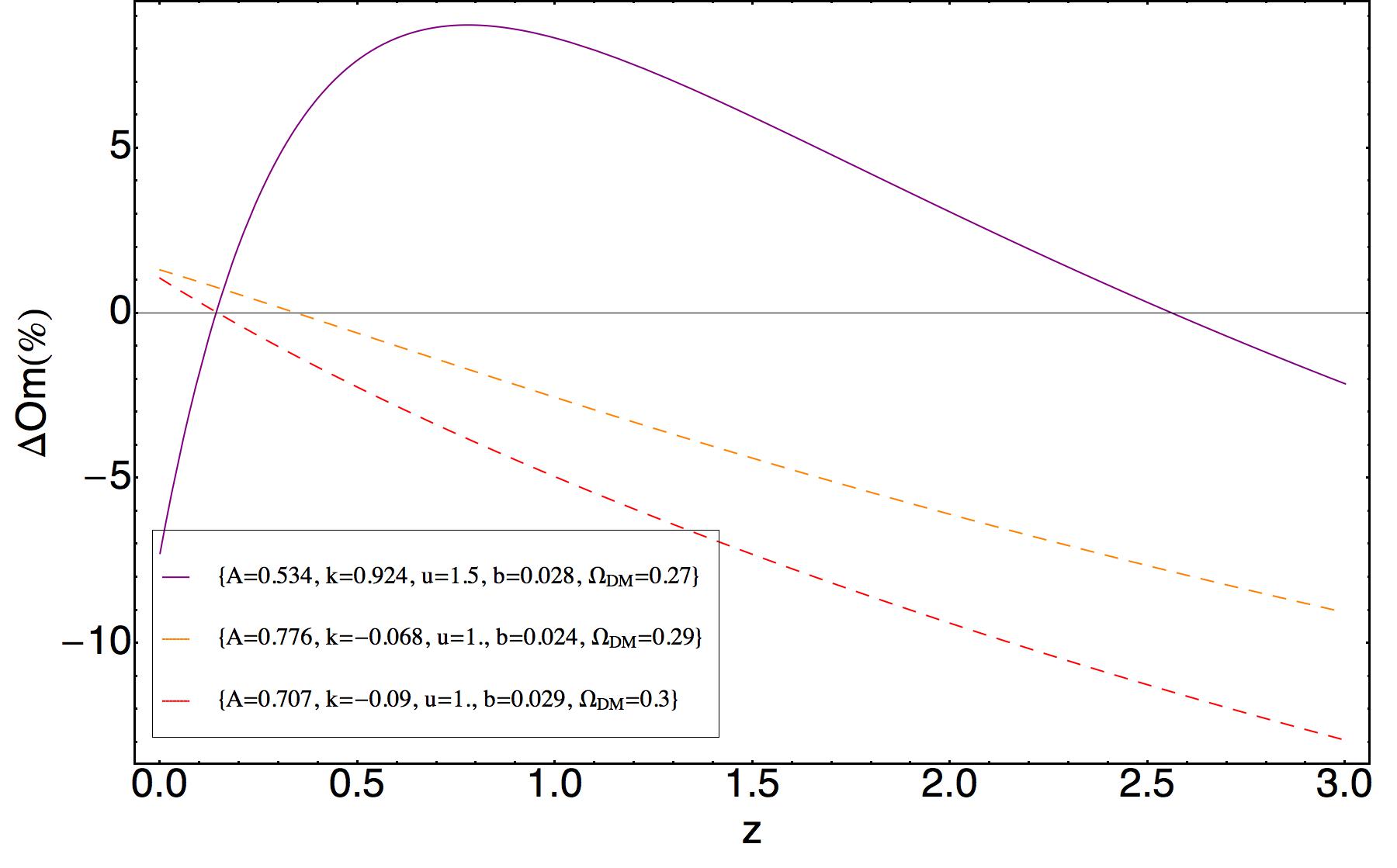} \\
  \end{array}$
  \end{center}
\caption{The graphical behavior of the deceleration parameter $q$ and 
$\Delta Om$, Eq.~(\ref{eq:DeltaOm}), against the redshift $z$. The 
considered model is free from the cosmological coincidence problem. The 
form of non-gravitational interaction is given by Eq.~(\ref{eq:M3}). The 
presented result corresponds to the analysis with $\chi^{2}_{OHD} + 
\chi^{2}_{BAO} + \chi^{2}_{\mu}$. }
  \label{fig:Fig4}
\end{figure}

The third model of this paper admits the following form of 
non-gravitational interaction
\begin{equation}\label{eq:M3}
Q = 3 H b \left( \rho_{de} + \rho_{dm} + \frac{\rho_{dm}^{2}}{\rho_{de} 
+ \rho_{dm}}\right),
\end{equation}
which gives the best fit of theoretical results with observational data 
when $\chi^{2}_{OHD} + \chi^{2}_{BAO} + \chi^{2}_{\mu} = 563.25$ and 
$H_{0} = 71.9$, $\Omega^{(0)}_{DM} = 0.27$, $u = 1.5$ are fixed in 
advance and the rest of the parameters of the model are defined as 
follows: $\{A, b, k \} = \{ 0.534, 0.028, 0.924 \}$. In this case we see 
that presented result coincidence with the result obtained for the 
model, when the non-gravitational interaction is given by 
Eq.~(\ref{eq:M2_3}). The consideration of the cases with $u = 1.0$ and 
$u = 1.25$~(with fixed $H_{0} = 71.9$ and $\Omega^{(0)}_{DM} = 0.27$) 
provided the best fit when $\{A, b, k \} = \{ 0.534, 0.028, 0.924 \}$ 
and $\{A, b, k \} = \{ 0.931, 0.014, -0.024 \}$, respectively. On the 
other hand, with $u = 1.0$~(with fixed $H_{0} = 71.9$ and 
$\Omega^{(0)}_{DM} = 0.29$) the best fit has been found with 
$\chi^{2}_{OHD} + \chi^{2}_{BAO} + \chi^{2}_{\mu} = 564.25$ when $\{A, 
b, k \} = \{ 0.776, 0.024, -0.068 \}$, while with fixed $H_{0} = 71.9$, 
$\Omega^{(0)}_{DM} = 0.30$ and $u = 1.0$ the best fit has been found 
when $\{A, b, k \} = \{ 0.707, 0.029, -0.09 \}$. The last state is 
described by $\chi^{2}_{OHD} + \chi^{2}_{BAO} + \chi^{2}_{\mu} = 
567.61$. For this model, the study of the question how the strong 
gravitational lensing data will affect on the best fit values of the 
parameters has been left as a topic of another study. Preliminary study 
presented here, showed that the best fit result also satisfying to the 
constraints imposed from BOSS and PLANCK 2015 experiments, should be 
accounted the result corresponding to the fixed $\Omega^{(0)}_{DM} = 
0.30$ case with $z_{tr} \approx 0.85$, $q \approx -0.58$ and $\Delta Om 
= 1\%$ at $z=0.0$. Moreover, as can be seen from the right plot of 
Fig.~(\ref{fig:Fig4}) $\Delta Om$ is an increasing function from the 
redshift.

\subsubsection{Case $1$}

In this section we will present the results of the fit for the 
phenomenological model, when the non-gravitational interaction is given 
in the following way
\begin{equation}\label{eq:M3_1}
Q = 3 H b \left( \rho_{de} + \frac{\rho_{dm}^{2}}{\rho_{de} + 
\rho_{dm}}\right),
\end{equation}
which is a particular case of more general form given by 
Eq.~(\ref{eq:M3}). First of all we would like to present the results 
according to $\chi^{2}_{OHD} + \chi^{2}_{BAO} + \chi^{2}_{\mu}$ 
constrain and  compare them with the results obtained from 
$\chi^{2}_{OHD} + \chi^{2}_{BAO} + \chi^{2}_{SGL} + \chi^{2}_{\mu}$ 
constrain. For instance, the study shows that when $H_{0} = 71.9$, 
$\Omega^{(0)}_{DM} = 0.27$ and $u = 1.5$ the minimal $\chi^{2}_{OHD} + 
\chi^{2}_{BAO} + \chi^{2}_{\mu} = 563.24$ will be obtained providing the 
best fit. In this case for the rest parameters we obtained $\{A, b, k \} 
= \{ 0.586, 0.047, 0.946 \}$. On the other hand, when $\Omega^{(0)}_{DM} 
= 0.28$, then the best fit with  $\{A, b, u, k \} = \{ 0.896, 0.047, 
1.25, 0.505 \}$~($\chi^{2}_{OHD} + \chi^{2}_{BAO} + \chi^{2}_{\mu} = 
562.70$) will be obtained. Moreover, the study showed that the model 
with $\Omega^{(0)}_{DM} = 0.29$, $\Omega^{(0)}_{DM} = 0.30$ and 
$\Omega^{(0)}_{DM} = 0.31$ provides the best fit when $\{A, b, u, k \} = 
\{ 0.896, 0.059, 1.25, 0.505 \}$~($\chi^{2}_{OHD} + \chi^{2}_{BAO} + 
\chi^{2}_{\mu} = 564.20$), $\{A, b, u, k \} = \{ 0.638, 0.047, 1.0, 
-0.112 \}$~($\chi^{2}_{OHD} + \chi^{2}_{BAO} + \chi^{2}_{\mu} = 567.63$) 
and $\{A, b, u, k \} = \{ 0.638, 0.057, 1.0, -0.112 \}$~($\chi^{2}_{OHD} 
+ \chi^{2}_{BAO} + \chi^{2}_{\mu} = 572.83$), respectively. Now, 
including strong gravitational lensing data, we obtained the following 
results. In particular, when $\Omega^{(0)}_{DM} = 0.27$ the best fit 
will be obtained when $\{A, b, u, k \} = \{ 0.586, 0.047, 1.5, 0.946 
\}$~($\chi^{2}_{OHD} + \chi^{2}_{BAO} + \chi^{2}_{SGL} + \chi^{2}_{\mu} 
= 781.21$) i.e. the best fit values of the parameters will not be 
affected. On the other hand, when $\Omega^{(0)}_{DM} = 0.28$, then the 
consideration of strong gravitational lensing data will significantly 
affect on the best fit values of the parameters - $\{A, b, u, k \} = \{ 
0.586, 0.057, 1.5, 0.946 \}$~($\chi^{2}_{OHD} + \chi^{2}_{BAO} + 
\chi^{2}_{SGL} + \chi^{2}_{\mu} = 780.64$). The results for 
$\Omega^{(0)}_{DM} = 0.29, 0.30, 0.31$ are presented in 
Table~\ref{tab:Table5} and it can be seen, that including of strong 
gravitational lensing data under the consideration will not affect the 
best fit values of the parameters. Future constraints mentioned earlier 
in this paper, support the case with $\Omega^{(0)}_{DM} = 0.30$ to be 
the candidate for the best fit. On the other hand, the results 
corresponding to $\Omega^{(0)}_{DM} = 0.31$ also can be counted to 
satisfy to imposed constraints. The left plot of Fig.~(\ref{fig:Fig4}) 
indicates how the relative change $\Delta Om$ evolves with the evolution 
of the universe.

\begin{table}
   \centering
     \begin{tabular}{ | l | l | l | l | l | l | p{1cm} |}
     \hline
  $\chi^{2}$ & $\Omega^{(0)}_{DM} (f)$ & $H_{0} (f)$ & $A$ & $b$ & $u 
(f)$ & $k$ \\
      \hline
  $782.11$ & $0.29$ & $71.9$ & $0.896$ & $0.057$ & $1.25$  & $0.505$ \\
      \hline
  $785.56$ & $0.30$ & $71.9$ & $0.638$ & $0.047$ & $1.0$  & $-0.112$ \\
      \hline
  $790.73$ & $0.31$ & $71.9$ & $0.638$ & $0.057$ & $1.0$  & $-0.012$ \\
      \hline
     \end{tabular}
\caption{The best fit results for the model with $Q = 3 H b \left( 
\rho_{de} + \frac{\rho_{dm}^{2}}{\rho_{de} + \rho_{dm}}\right)$ with 
$\chi^{2}_{OHD} + \chi^{2}_{BAO} + \chi^{2}_{SGL} + \chi^{2}_{\mu}$. $f$ 
means that the parameter has been fixed to the presented value in 
advance before the fit has been started.}
   \label{tab:Table5}
\end{table}

\begin{figure}[h!]
  \begin{center}$
  \begin{array}{cccc}
\includegraphics[width=80 mm]{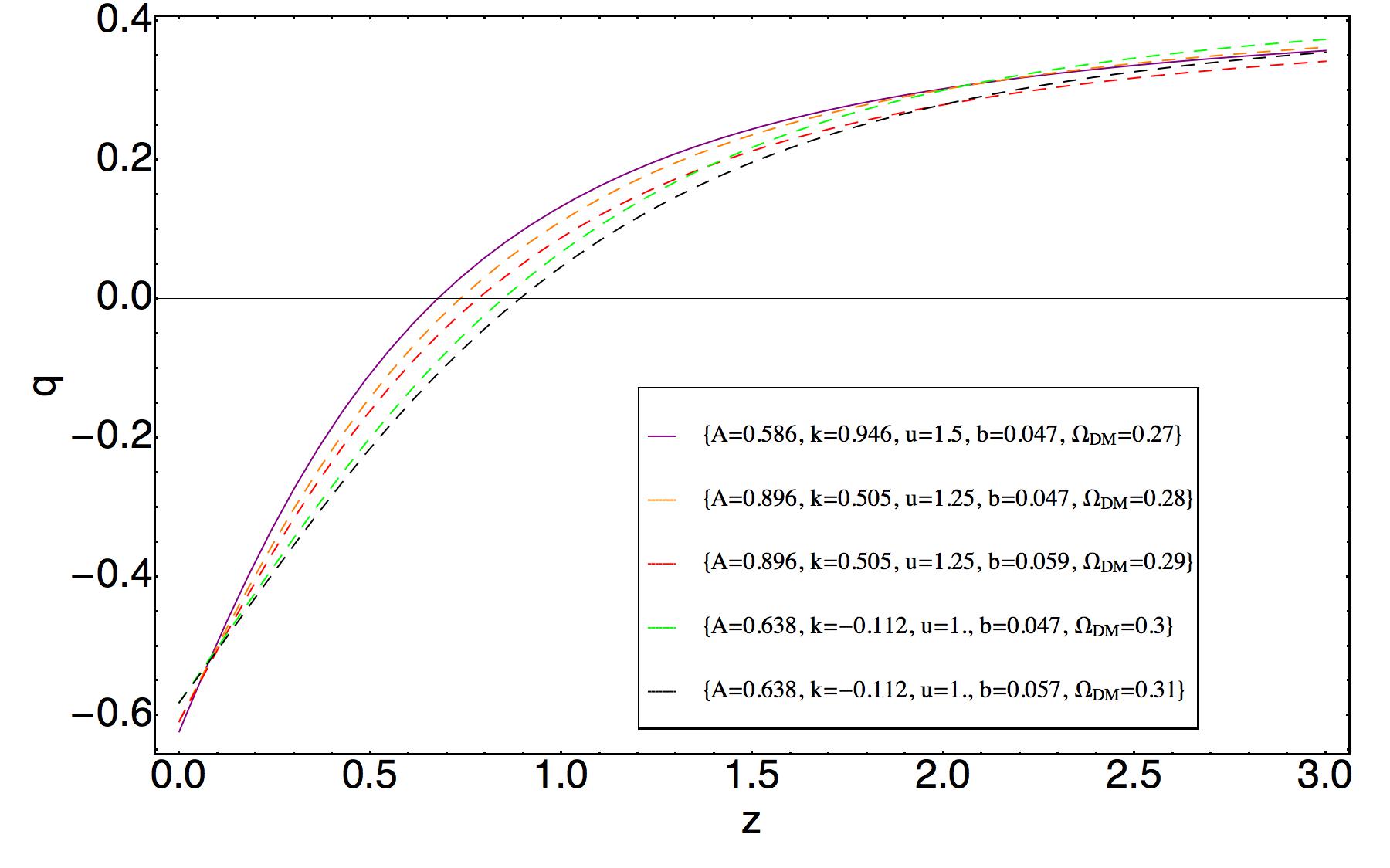} &
\includegraphics[width=80 mm]{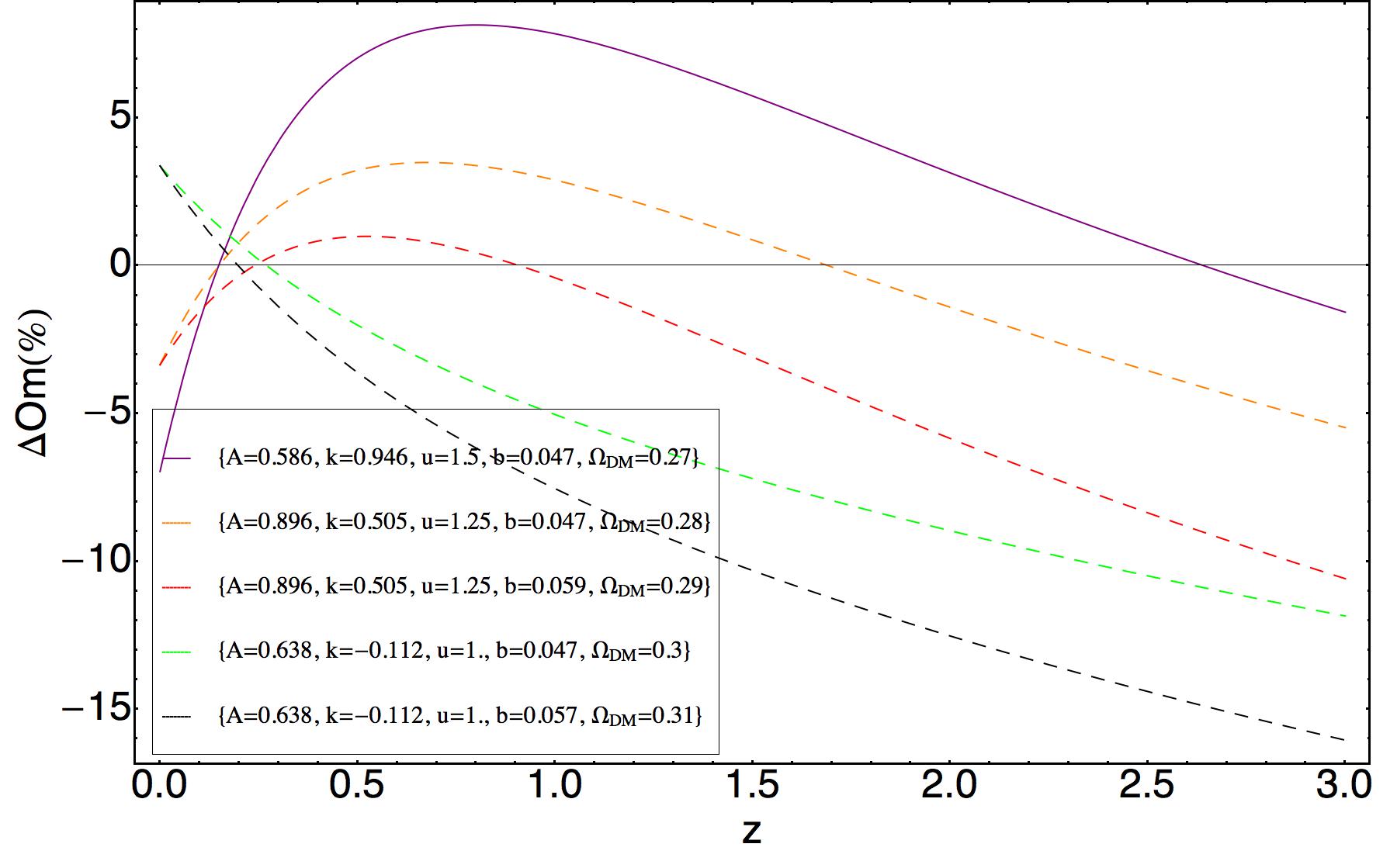} \\
  \end{array}$
  \end{center}
\caption{The graphical behavior of the deceleration parameter $q$ and 
$\Delta Om$, Eq.~(\ref{eq:DeltaOm}), against the redshift $z$. The 
considered model is free from the cosmological coincidence problem. The 
form of non-gravitational interaction is given by Eq.~(\ref{eq:M3_1}). 
The presented result corresponds to the analysis with $\chi^{2}_{OHD} + 
\chi^{2}_{BAO} + \chi^{2}_{\mu}$.}
  \label{fig:Fig5}
\end{figure}

\subsubsection{Case $2$}

\begin{figure}[h!]
  \begin{center}$
  \begin{array}{cccc}
\includegraphics[width=80 mm]{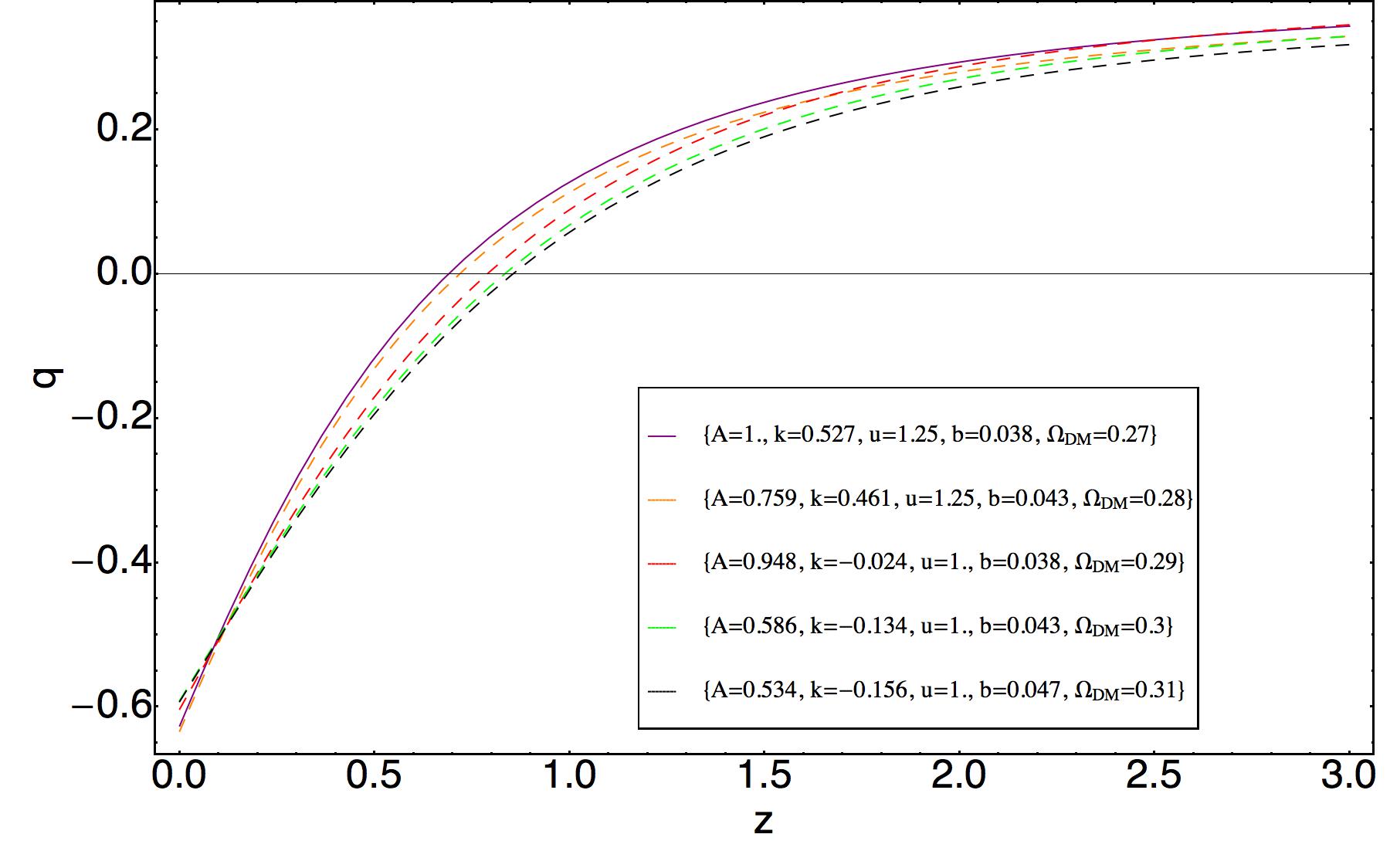} &
\includegraphics[width=80 mm]{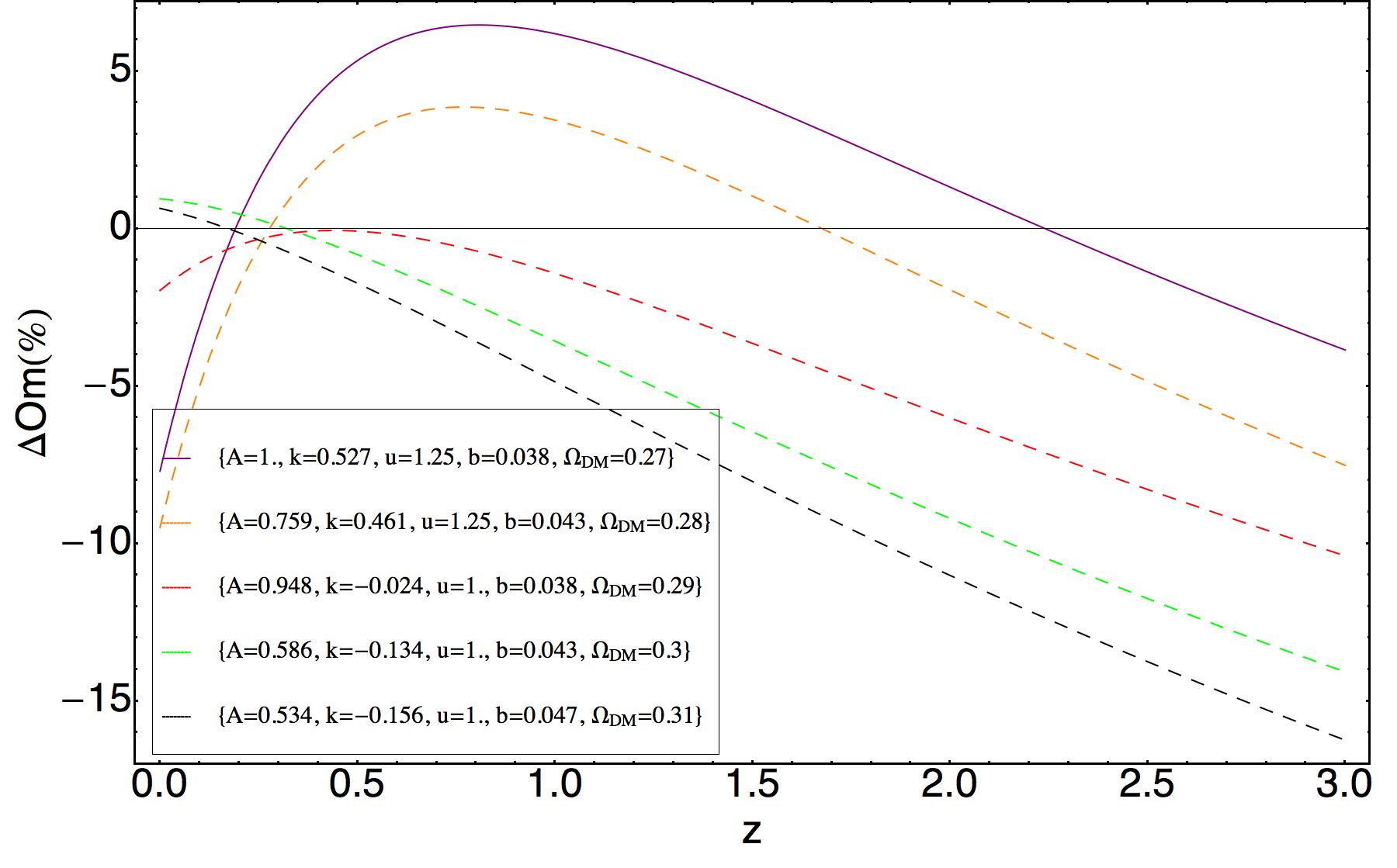} \\
  \end{array}$
  \end{center}
\caption{The graphical behavior of the deceleration parameter $q$ and 
$\Delta Om$, Eq.~(\ref{eq:DeltaOm}), against the redshift $z$. The 
considered model is free from the cosmological coincidence problem. The 
form of non-gravitational interaction is given by Eq.~(\ref{eq:M3_2}).}
  \label{fig:Fig6}
\end{figure}

The last model studied in this work admits the following form of 
non-gravitational interaction
\begin{equation}\label{eq:M3_2}
Q = 3 H b \left( \rho_{dm} + \frac{\rho_{dm}^{2}}{\rho_{de} + 
\rho_{dm}}\right).
\end{equation}
During the study of the model, when in addition to $\chi^{2}$ analysis 
with $\chi^{2}_{OHD} + \chi^{2}_{BAO} + \chi^{2}_{SGL} + 
\chi^{2}_{\mu}$, we applied the constraints from BOSS and PLANCK 2015 
experiments, and take into account the constrains from modified 
two-point $Om$ analysis gives us the best fit values of the parameters 
of the model as follows: $\omega_{de} \approx -1.039$ at $z=0.0$ with 
$\{ H_{0}, \Omega^{(0)}_{dm}, A, b, u, k \} = \{ 71.9, 0.30, 0.638, 
0.047, 1.0, -0.112 \}$. The graphical behavior of the deceleration 
parameter is presented on the left plot of Fig.~(\ref{fig:Fig6}), while 
the graphical behavior of $\Delta Om$ is presented on the right plot. 
Both clearly indicates how the mentioned parameters evolve during the 
evolution of the universe, moreover, it is possible also to estimate the 
present day values of them very easily. During the study of the behavior 
of the equation of state parameter of considered polytropic fluid for 
all models we found two possibilities. In particular we observed that 
for some models~(the difference between the models is the form of 
non-gravitational interaction) $\omega_{de} > 0$ at high redshifts and 
there is a phase transition to a phantom dark fluid state satisfying to 
the constraints on $\omega_{de}$ according to PLANCK 2015 experiment. 
However, there is also possibility to have a phantom - phantom 
transitions also providing the accelerated expansion of the universe. We 
would like to mention that mentioned phantom - phantom transitions have 
been observed firstly in scope of generalized holographic dark energy 
models with Nojiri-Odintsov cut-offs.

\begin{table}
   \centering
     \begin{tabular}{ | l | l | l | l | l | l | p{1cm} |}
     \hline
  $\chi^{2}$ & $\Omega^{(0)}_{DM} (f)$ & $H_{0} (f)$ & $A$ & $b$ & $u 
(f)$ & $k$ \\
      \hline
  $781.17$ & $0.27$ & $71.9$ & $1.0$ & $0.038$ & $1.25$  & $0.527$ \\
      \hline
  $780.59$ & $0.28$ & $71.9$ & $0.759$ & $0.043$ & $1.25$  & $0.461$ \\
      \hline
  $782.10$ & $0.29$ & $71.9$ & $0.948$ & $0.038$ & $1.0$  & $-0.024$ \\
      \hline
  $785.51$ & $0.30$ & $71.9$ & $0.586$ & $0.043$ & $1.0$  & $-0.134$ \\
      \hline
  $790.75$ & $0.31$ & $71.9$ & $0.534$ & $0.047$ & $1.0$  & $-0.156$ \\
      \hline
     \end{tabular}
\caption{The best fit results for the model with $Q = 3 H b \left( 
\rho_{dm} + \frac{\rho_{dm}^{2}}{\rho_{de} + \rho_{dm}}\right)$ with 
$\chi^{2}_{OHD} + \chi^{2}_{BAO} + \chi^{2}_{SGL} + \chi^{2}_{\mu}$. $f$ 
means that the parameter has been fixed to the presented value in 
advance before the fit has been started.}
   \label{tab:Table7}
\end{table}

\section{\large{Discussion}}\label{sec:Discussion}

Available observational data suggests to include dark energy and dark 
matter in general relativity to explain the accelerated expansion of the 
universe. On the other hand, it is possible to introduce 
non-gravitationally interacting dark energy and dark matter to solve the 
problems of the large scale universe. The duality is on face with its 
problematic consequences. From one hand side, seems that it will be easy 
to parameterize the dark side of the universe and solve the problems, 
but from the other hand side, it appears that existing tension between 
observational datasets makes additional complexity. Therefore, there is 
ongoing active research in order to find a solution involving many 
phenomenological assumptions. In particular, there are different 
phenomenological assumptions concerning to the form of dark energy and 
non-gravitational interaction. Motivated by recent developments, in this 
paper we considered new cosmological models involving new interacting 
varying polytropic gas models. In order to obtain the best fit values of 
the parameters of the models we used $\chi^{2}$ analysis involving the 
differential age of old galaxies, given by $H(z)$, the peak position of 
baryonic acoustic oscillations known as BAO data, the SN Ia data and 
strong gravitation lensing data. To simplify the analysis we fixed the 
values of some of the parameters before the fit has been start and kept 
them frozen in future, we involved constraints on the equation of state 
parameter of dark fluid $\omega_{de}$ from PLANCK 2015 experiment, then 
we took into account reported value for the Hubble parameter at $z = 
2.34$ from BOSS experiment. Moreover, we used constraints obtained from 
a modified two-point $Om$ analysis giving $Omh^{2}(z_{1};z_{2}) = 0.124 
\pm 0.045$,
$Omh^{2}(z_{1};z_{3}) = 0.122 \pm 0.01$ and $Omh^{2}(z_{2};z_{3}) = 
0.122 \pm 0.012$ for $z_{1} = 0$, $z_{2} = 0.57$ and $z_{3} = 2.34$, 
respectively~\cite{SahniFin}. Mentioned additional constrains allowed us 
to establish the best fit values of the parameters of the models with 
fixed $H_{0}$, $u$ and $\Omega^{(0)}_{DM}$. The study shows, that 
considered models, which are differ from each other by the form of 
non-linear non-gravitational interactions between dark energy and dark 
matter can explain the accelerated expansion. It is possible to explain 
the phase transition between decelerated expanding and accelerated 
expanding phases during the evolution of the universe. Moreover, the 
study of the relative change of $Om$ parameter shows clear difference 
between new models and $\Lambda$CDM standard model of cosmology. The 
main interesting result has been observed during the study of the 
equation of state parameter of polytropic fluid. In particular, the 
study shows that considered models when non-gravitational interactions 
are given by Eq.~(\ref{eq:M1_1}), Eq.~(\ref{eq:M1_2}), 
Eq.~(\ref{eq:M1}), Eq.~(\ref{eq:M2_1}), respectively, then quintessence 
- phantom transition for the equation of state parameter of dark energy 
will be observed. On thi other hand, when we consider non-gravitational 
interactions given by Eq.~(\ref{eq:M2_2}), Eq.~(\ref{eq:M2_3}), 
Eq.~(\ref{eq:M3}), Eq.~(\ref{eq:M3_2}), then phantom - phantom phases 
unification will be observed. However, when the interaction, for 
instance, is given by Eq.~(\ref{eq:M1_2}), then the results 
corresponding to $\Omega^{(0)}_{DM} = 0.31$ satisfying the the 
considered constraints, provides dark energy with phantom - phantom 
transition. On the other hand, when the interaction is given by 
Eq.~(\ref{eq:M1_2}), then we will observe also quintessence - phantom 
transition. In summary - we need more observational data in order to be 
able to choose the best model of interacting varying politropic dark 
fluid model from the models considered in this paper. It can be done, 
for instance, involving constraints provided by the study of the 
structure formation. Moreover, this will allow to define which one of 
mentioned transitions for dark energy is the realistic scenario, because 
from the study of the deceleration parameter and the equation of state 
parameter this question cannot be answered. If it will be found that a 
model with phantom - phantom transition will be the best model among 
considered models, then it is necessary to find Nojiri-Odintsov 
holographic dark energy representation of the model and study the 
application of the model to the inflationary expansion phase of the 
universe. In future research reconstruction of modified theories of 
gravity for considered models should be performed and since we saw a 
non-unique imprint of the type of non-gravitational interaction on the 
behavior of the equation of state parameter of dark fluid, then it is 
necessary to determine the type of future singularities which will 
provide additional sources to extend applied constraints used in this 
paper. Mentioned possibilities and tasks are the subject of additional 
research and will be reported in another paper.

\section*{\large{Acknowledgments:}}
Martiros Khurshudyan appreciates Prof. K. Urbanowski from Institute of 
Physics, University of Zielona Gora, for valuable comments and 
suggestions during the preparation of the paper.

\section*{\large{Author Contributions:}}
Martiros Khurshudyan designed the main subject of this research and 
mainly wrote the manuscript.
Asatur Khurshudyan commented on the manuscript at all stages with 
performing part of the analysis. All authors equally promoted the 
research and discussed the results. All authors have read and approved 
the final manuscript.

\section*{\large{Conflicts of Interest:}} The authors declare no 
conflict of interest.

\end{document}